\documentclass[12pt]{article}

\usepackage[american]{babel}
\usepackage[dvips]{graphicx}
\usepackage{amssymb}
\usepackage{amsmath}
\usepackage{color}

\newcommand{\singlespace}{\renewcommand{\baselinestretch}{1}\large\normalsize}
\newcommand{\doublespace}{\renewcommand{\baselinestretch}{1.6}\large\normalsize}
\newcommand{\bra}[1]{\langle {#1} \vert}
\newcommand{\ket}[1]{\vert {#1} \rangle}
\newcommand{\Kbar}{\overline{K}}
\newcommand{\beq}{\begin{equation}}
\newcommand{\eeq}{\end{equation}}
\newcommand{\KObar}{\overline{K^0}}

\begin{document}

\selectlanguage{american}

\begin{center}
\doublespace 
\begin{Large}
{\bf  Medium modifications of antikaons in dense matter} 
\end{Large}
\vskip 0.4in
Th. Roth$^a$, M. Buballa$^a$, and J. Wambach$^{a,b}$
\\[4mm]
{\small{\it $^a$ Institut f\"ur Kernphysik, TU Darmstadt,\\
            Schlossgartenstr. 9, D-64289 Darmstadt, Germany\\[1mm]
             $^b$ Gesellschaft f\"ur Schwerionenforschung (GSI),\\
             Planckstr. 1, D-64291 Darmstadt, Germany}}
\end{center}

\vspace{8mm}

\begin{abstract}
We investigate the modification of antikaons in isospin symmetric and 
asymmetric nuclear matter.
Using the leading s--wave couplings of the SU(3) chiral meson--baryon
Lagrangian we solve the coupled channel kaon--nucleon scattering equation 
selfconsistently.
The in--medium antikaon propagator is calculated for different densities and 
different proton/neutron ratios. The spectral function of the antikaon is 
found to be broadened strongly, and its mass is shifted downward significantly.
However, comparing the effective in--medium mass of the antikaon to the
relevant charge chemical potential of neutron star matter, we
find  that kaon condensation is unlikely to occur in neutron stars.
\end{abstract}

\singlespace

%%=============================================================

\section{Introduction}

Understanding the properties of strongly interacting matter under extreme
conditions such as high temperatures or densities is of considerable interest in hadronic physics.
In this context strange particles play a key role. 
In ultrarelativistic heavy--ion collisions, primarily exploring the 
high--temperature regime of the QCD phase diagram, strangeness enhancement
is considered to be one of the possible signals for the formation of
the quark--gluon plasma \cite{RaMu}. 
At low temperatures but high densities, as present, e.g., in the interiors
of neutron stars, strangeness might show up in the form of a condensate of
negative kaons. This possibility was first pointed out by Kaplan and 
Nelson \cite{KapNel,KapNel2} and has received great attention since. 
The appearance of an antikaon condensate would soften the equation of state of 
the star, allowing more compact stellar objects. 

The basic considerations leading to this idea are relatively 
simple \cite{Koch3}:
Since kaons, as the lightest mesons with strangeness, are the 
(pseudo--) Goldstone bosons of $SU(3)$ chiral symmetry breaking,
their interactions can be studied within chiral perturbation theory
($\chi PT$). At lowest order, the corresponding kaon--nucleon
scattering Lagrangian contains two s--wave interaction terms:
a constant scalar term due to explicit chiral symmetry breaking 
and a momentum--dependent vector term, the so--called
Weinberg--Tomozawa term \cite{WeinbergTomo,WeinbergTomo1}. 
The scalar interaction is attractive for both, kaons and antikaons,
while the vector interaction is repulsive for kaons ($K^+$ and $K^0$)
but attractive for antikaons ($K^-$ and $\bar{K^0}$).
As a result, one expects an interaction which at tree level
is weakly repulsive for kaons but strongly attractive for antikaons. 
Hence, despite  large quantitative uncertainties, the antikaon mass
should decrease in matter, eventually leading to kaon condensation,
once the mass drops below the chemical potential of the surrounding leptons.

Unfortunately, the situation is not that simple. 
Firstly, it should be noted that there is an ongoing controversial discussion
about possible ambiguities related to the off--shell extrapolation from the 
$\Kbar N$--scattering amplitude to the kaon selfenergy
 \cite{Myhrer93,Myhrer94,ThorssonWirzba,WirzbaThorHirsch}.
Moreover, the empirical value of the s--wave $K^-$ nucleon 
scattering length is {\it repulsive} at threshold 
(${\Re} e\, a_{K^- p} = -0.78$ fm).
This discrepancy between experiment and the above considerations can be 
attributed to the existence of the $\Lambda(1405)$ resonance 
just below the $KN$ threshold, which gives rise to a repulsive contribution to the 
scattering amplitude at threshold. Apart from a few exceptions where the $\Lambda(1405)$ 
has been introduced as an elementary field \cite{Krippa}, in most models it is 
dynamically generated through the $\Kbar N$ -- scattering process 
\cite{Koch,NuclPhysA612,Lutz,Oset}. In this case it 
cannot be treated perturbatively. 

On the other hand, there are various experimental indications that antikaons,
once they propagate through dense nuclear matter, indeed feel an attractive 
potential.
Studies of the energy shifts and the widths of the lowest
levels in kaonic atoms probe the behavior of antikaons in 
nuclear matter of densities between zero and $\rho_0$.
A collection of data over a wide range of atoms was analyzed by
Friedman \cite{Fried}. These results require an attractive potential
for the antikaons, although different theoretical approaches 
lead to quite different values for the depth of this 
potential  \cite{Galetal,Tolos0007042}.

This finding is supported by the kaon data in heavy--ion collisions. 
The azimuthal emission patterns of $K^+$ and $K^-$ obtained 
by the FOPI and KaoS collaborations \cite{FOPI,KAOS} indicate a
different in--medium behavior for kaons and antikaons, corresponding
to repulsive and attractive interactions, respectively.
Naively, this also gives a simple explanation for the enhanced 
$K^-/K^+$ ratios which are found in nucleus--nucleus collisions as compared 
with (anti--) kaon production in $NN$:
Lowering the $K^-$ mass would reduce its production threshold and thus
increase the yield \cite{Hirschegg2000,Hirschegg2000Devismes,Hirschegg2001}.
However, the situation is complicated by the fact that in dense matter 
the $K^-$ yield is enhanced, even without medium modification of
its mass. This is because there are hyperons around (e.g., owing to prior 
$K^+$ production) which enable secondary production mechanisms,
as $\pi Y \rightarrow K^- N$.
Hence, for a consistent interpretation of the data, we do not only need 
a good understanding of the in--medium modification of the (anti--) kaons,
but also of the above production process. 

It is exactly this coupling of 
$\pi Y$ to $\Kbar N$ that turns out to be the key to the interaction of the 
$\Kbar$ with matter, giving rise to the $\Lambda(1405)$ resonance. This
situation requires the treatment in terms of coupled channels, leading to 
the $\Lambda(1405)$ resonance and connected effects.
Performing the coupled channel calculation at finite density results in a $\Lambda(1405)$
that is shifted upwards in energy \cite{Koch,NuclPhysA612,Weise,Schaffner}. 
At lower energies around the original (vacuum)  threshold the optical 
potential for the $K^- $ becomes attractive, which would explain the data 
from kaonic atoms. 

Still, one has to be careful with this kind of analysis. In the first works
that went beyond a mean field description to include the
$\Lambda(1405)$ (e.g.~\cite{Koch,NuclPhysA612,Weise}),
the repercussion of the dynamics of the $\Lambda(1405)$ on the
antikaons were not included selfconsistently. However, this turned out to
be of great importance: The selfconsistent $\Lambda(1405)$ does not move
very much \cite{Galetal,Lutz,LutzKorpa}, rendering the argument for the kaonic atoms 
questionable.
 Obviously, a comprehensive treatment of all the effects becomes
necessary. This is what we aim at in this work.
To this end, we basically carry out the following program,
which is sketched in Fig.~\ref{overview_fig}:

%%%%%%%%%%%%%%%%%%%%%%%%%%%%%%%%%%%%%%%%%%%%%%%%%%%%%%%%%%%%%
\begin{figure}[htp]
\begin{center}
\includegraphics[width=8cm]{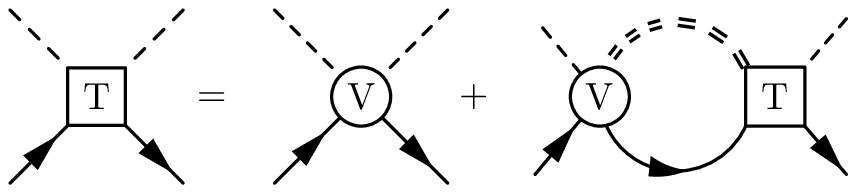}

\includegraphics[width=8cm]{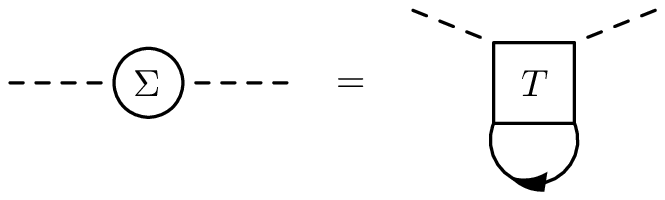}

\includegraphics[width=10cm]{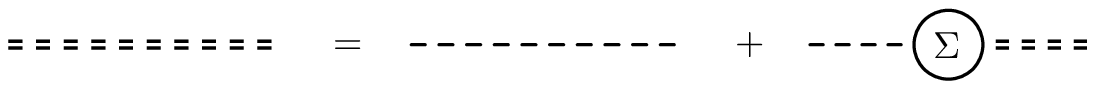}
\caption[$\Kbar N -- T$ matrix and antikaon selfenergy]
{\small
 Illustration of the selfconsistency scheme:
 Scattering equation for the  antikaon--nucleon $T$ matrix (upper line),
 in--medium antikaon self--energy (center), and dressed kaon
 propagator (lower line).} 
\label{overview_fig}
\end{center}
\end{figure}
%%%%%%%%%%%%%%%%%%%%%%%%%%%%%%%%%%%%%%%%%%%%%%%%%%%%%%%%%%%%%

In a first step we construct the $\Kbar N$ T--matrix, taking into account
the coupling of $\Kbar N$ to all other important channels.  
Here the Lagrangian of chiral perturbation theory  provides the basic
interaction, i.e., the scattering kernel 
for the various mesons and baryons we take into account.
As already indicated, the $\Lambda(1405)$ appears slightly 
(about 30~MeV) below the $\Kbar N$
threshold which prohibits a purely perturbative treatment of the scattering
process. Instead, we take the simplest diagram formed from 
${\cal L}_{\chi PT}$ as the kernel of the Bethe--Salpeter scattering 
equation,  sum it to all orders and get the $\Lambda(1405)$   
as a dynamically generated resonance. 
In this context the coupling to other meson--baryon channels,
in particular the  $\pi \Sigma$ channel, which opens about 70~MeV below
the resonance, is crucial. 

Having constructed the scattering amplitudes in vacuum, 
we take the calculation to the dense medium,
replacing the vacuum propagators by in--medium ones.  
The in--medium self--energy of the antikaon can be obtained 
from the $T$--matrix by closing a nucleon (hole) loop 
(see Fig.~\ref{overview_fig}).  
This leads to a modified kaon propagator which, in turn, should be
used to calculate the $T$--matrix.  
This procedure has to be iterated until a selfconsistent result is achieved.

A very important aspect in our calculations is the
consideration of the momentum dependence of all calculated quantities. 
Calculating
the antikaon selfenergy from the in--medium scattering $T$--matrix means an
integration over momentum, so the $T$--matrix, depending on energy and momentum
separately in the medium, has to be known as a function of energy and
momentum. This requirement becomes even stronger once we iterate the procedure
to selfconsistency.   

This is not just a technical issue. In general the existence of the nuclear
matter rest frame as a preferred frame of reference forces us to
specify explicitly the energy and momentum dependence of in--medium properties.    
In addition, considering  heavy--ion collisions, the optical potential of 
the antikaon 
is probed at finite momentum. At  a moderate temperature of $T=80$
MeV \cite{FOPI} the antikaon has an average momentum of more than $300$~MeV with 
respect to
the matter rest frame. Momenta  above $\sim 300$ MeV are typical in
the experimental data on heavy ion collisions. Nevertheless, in many  
calculations it has been assumed that the attractive potential for the
antikaons, i.e.~their change in the mass is independent of momentum 
\cite{NuclPhysA612,Oset-7_99}. This is not justified. To the
contrary, a considerable momentum dependence of all in--medium
properties is found \cite{Schaffner}, see also section 
\ref{subsecMomentumdependence}.

In this article we focus on $s$--wave interactions. 
Kaon condensation is thus connected to the propagator at zero momentum.
$p$--wave antikaon condensation has been investigated elsewhere
\cite{LiLeeBrown,Kcond-BrownRapp,KolVos,KolVosKaemp,Muto,KolVos2}. 
One aspect here is the effect of hyperon--nucleon--hole contributions 
on the antikaon selfenergy. 
We briefly investigate these contributions in Sect.~\ref{pwavekaonselfenergy}. 
As they turn out to be very small, but would lead to considerable technical 
complications,
we carry out the selfconsistency problem with the $s$--wave part 
alone.

The remainder of this work is organized as follows. 
In Sect.~\ref{Detailed KN scattering}, we briefly discuss  
the formalism involved in calculating the $SU(3)$ meson--baryon scattering 
in a coupled channels scheme, based on the $\chi PT$ Lagrangian.  
After that, in Sect.~\ref{Kaon_selfenergy}, the model is used to investigate 
the behavior of antikaons in dense isospin symmetric matter. 
Among other aspects we study the effect of the selfconsistent treatment, 
the momentum and density dependence of the propagator, and the role
of the in--medium modification of pions.
In Sect.~\ref{chapAsymm} the analysis is extended to isospin asymmetric 
matter, i.e., to systems with different densities of protons and neutrons, 
as found in realistic systems. The implications of these results for 
the question of antikaon condensation in neutron stars are discussed 
Sect.~\ref{chapKcondens}. Our main results are summarized in 
Sect.~\ref{cSummary}.

%%=============================================================

\section{Meson--baryon scattering}
\label{Detailed KN scattering}

As outlined above (see Fig.~\ref{overview_fig}),
the first step in our procedure to construct the 
in--medium kaon propagator consists in a coupled--channel calculation of
the $\bar K N \rightarrow \bar K N$ scattering amplitude,
\beq
\bra{\overline{K}(k')\,N(p')}\,T\,\ket{\overline{K}(k)\,N(p)}\,=\,
(2\pi)^4\,\delta^4(k'+p'-k-p)\,\overline{u}(p')T_{KN\rightarrow KN}
(k',p';k,p)\,u(p)~.
\label{Tmatrixonshell}
\eeq
The $T$ matrix elements are given by the Bethe--Salpeter equation \cite{BS}, 
\beq
    T_{fi} \;=\; V_{fi} \;+\;\sum_c \int \frac{d^4 l}{(2\pi)^4}
    V_{fc}\,{\cal G}^{\scriptscriptstyle BS}_{c}\,T_{ci}~,
\label{BSE}
\eeq    
which is graphically sketched in the upper line of Fig.~\ref{overview_fig}. 
We will refer to this equation also as the $T$--matrix equation. 
The indices $i$ and $f$ denote the incoming and outgoing meson--baryon
channels, while the index $c$ indicates the intermediate states, i.e., 
the meson--baryon channels that form the loop. 
For brevity we have suppressed all momentum arguments and only kept
the loop momentum $l$.

Input to the $T$--matrix equation are the interaction kernel $V$,
which will be specified in Sect.~\ref{interaction}, 
and the two--particle propagator ${\cal G}^{\scriptscriptstyle BS}$.
The latter is basically the product of the single--particle propagators
of the baryon $B_c$ and the meson $M_c$ in the loop,
${\cal G}^{\scriptscriptstyle BS}_{c} =i\,S_{B_c} G_{M_c}$.
In this work we include all channels of the SU(3) meson and baryon octets 
coupled to total strangeness $-1$, dropping the $\Xi$--channels that 
open only at larger energies.

%%------------------------------------------------------------------------------------------------

\subsection{Interaction}
\label{interaction}

As discussed earlier, the interaction kernel $V$ which
enters the $T$--matrix equation is derived from the $SU(3)$ chiral 
Lagrangian. The basic building blocks are the matrices 
\beq
    u \;=\;\exp{\Big(\frac{i\Phi}{f_\pi}\Big)}~,\qquad
    \Phi=\frac{1}{\sqrt{2}} 
    \begin{pmatrix} 
       \frac{\eta}{\sqrt{6}}+\frac{\pi^0}{\sqrt{2}}   & \pi^+   & K^+ \cr 
       \pi^- & \frac{\eta}{\sqrt{6}}-\frac{\pi^0}{\sqrt{2}}     & K^0 \cr 
       K^-   & \overline{K^0}          & -\frac{2\eta}{\sqrt{6}}
    \end{pmatrix}~,
\eeq
and
\beq
    B  = 
    \begin{pmatrix} 
      \frac{1}{\sqrt{6}}\Lambda+\frac{1}{\sqrt{2}}\Sigma^0  & \Sigma^+ & p \cr 
      \Sigma^- & \frac{1}{\sqrt{6}}\Lambda- \frac{1}{\sqrt{2}}\Sigma^0 & n \cr
      \Xi^-   &   \Xi^0    &   -\sqrt{\frac{2}{3}}\Lambda 
    \end{pmatrix}~,
\eeq
corresponding to the pseudoscalar meson and baryon octets, respectively.

At lowest chiral order, the Lagrangian contains two meson--baryon
interaction terms:
\beq
{\cal L}_{MB}^{(1)} \;=\;{\cal L}_{WT} \;+\;{\cal L}_{pw}~. 
\label{LMB1}
\eeq
The first term,
\beq
    {\cal L}_{WT} \;=\; i\,tr(\overline{B} \gamma^\mu [\Gamma_\mu, B])~,
    \qquad \text{with} \quad     
    \Gamma_\mu = \frac{1}{2} 
    (u^\dagger \partial_\mu u + u \partial_\mu u^\dagger)~,\quad
\label{WT}
\eeq
is the already mentioned Weinberg--Tomozawa term, which corresponds 
to $s$--wave interaction. 
As pointed out in the Introduction, this term is crucial for a realistic 
description of $\Kbar N$ scattering
and will give a main contribution to the interaction kernel.

The second term, ${\cal L}_{pw}$, corresponds
to $p$--wave interactions. As we will show in Sect.~\ref{pwavekaonselfenergy}~, it has only little 
effect in the $\Kbar N$ sector. 
Of course, $p$--wave interactions must be included to describe 
medium modifications of {\it pions}. 
However, since pions are not selfconsistently coupled to the kaon sector 
in our model, they can be treated separately.
This will be briefly discussed in Sect.~\ref{pions}.

At second chiral order there are three interaction terms 
due to explicit chiral symmetry breaking through the quark mass matrix
${\cal M}=\mathit{diag}(m_u,m_d,m_s)$.
These are the "Sigma terms" 
\cite{chiMei-5_02,Sigmaterms-Borasoy,BorasoyMeissner}, 
which are given by
\begin{equation}
\begin{split}
{\cal L}_{\Sigma} & = b_D\,tr(\overline{B} \{\chi_+,B\})\;
+\,b_F\,tr(\overline{B} [\chi_+,B])\;+\,b_o\,tr(\overline{B} B)\,tr(\chi_+)~, 
\end{split}
\label{Ls}
\end{equation}
with
\beq 
    \chi_+ \;=\; 2B_0(u^\dagger\,{\cal M}\,u^\dagger \,+\, u\,{\cal M}\,u)~.
\eeq
Although of higher chiral order, they are comparable in strength to the 
Weinberg--Tomozawa contribution when evaluated at
the respective meson--baryon thresholds. Therefore they should be taken into account.

While $B_0$ can be related to the chiral condensate via the 
Gell--Mann--Oakes--Renner relation, 
$B_0 = - \langle \bar qq \rangle/f_\pi^2$,
the constants $b_D, b_F $ and $b_0$ are examples of the so called
low--energy--constants (LECs). These parameters are not constrained by 
symmetries and have to be fixed from phenomenology. One can use the 
predictions of the theory for the baryon masses to pin down the values 
of at least $b_D$ and $b_F$. The values used for the calculations in 
this work  are taken from \cite{MFML}:
\beq
b_D\,=\,0.061\; \mbox{GeV}^{-1},\qquad b_F\,=\,0.195 \; \mbox{GeV}^{-1},
 \qquad b_0\,=\,-0.346\;  \mbox{GeV}^{-1}~.
\eeq

%%---------------------------------------------------------------------------------------------

\subsection{Factorization of the $T$--matrix equation}

The difficult point in solving the Bethe--Salpeter equation, 
Eq.~(\ref{BSE}), is any momentum dependence in the vertices. 
For constant interaction kernels $V$, the equation can be
factorized and reduces to a simple matrix equation which can 
straightforwardly be solved by inversion:
\beq
    T \;=\; V \;+\; V\,J\,T  \quad \Rightarrow \quad
    T \;=\; (1-V\,J)^{-1}\,V~.
\label{factor}
\eeq
Here $J$ is a diagonal matrix in the space of meson--baryon channels $c$
corresponding to the loop integrals    
\beq
J_c \;=\; i\,\int \frac{d^4 l}{(2\pi)^4} \, S_{B_c} \,G_{M_c}~.
\eeq
It depends on the total 4--momentum of the pair only and, hence,
as long as $V$ is constant, the $T$ matrix also depends only on the 
total 4--momentum of the meson--baryon system. 

However, in our case, $V$ is not constant, because the 
Weinberg--Tomozawa term, Eq.~(\ref{WT}), leads to 4--momentum dependent 
vertices via the derivatives $\partial_\mu$. 
To avoid this difficulty we follow the commonly used strategy which involves the 
following approximations: 
First, we evaluate ${\cal L}_{WT}$ in the baryon rest frame \cite{MFML,Oset}. 
This amounts to replacing
\beq
    {\cal L}_{WT}  \;\rightarrow\;  i\,tr(B^\dagger \,[\Gamma^0,B])\;,
\eeq
which removes the 3--momentum dependence. 
Still, the vertices depend on the energies $k^0$ and ${k'}^0$ of the
incoming and outgoing meson fields, respectively, i.e. 
\beq
    V_{WT} \;\propto\; \frac{i}{2 f_\pi}\,(k^0 +{k'}^0)~. 
\eeq
Therefore we solve the $T$--matrix equation by applying the so--called
on--shell factorization \cite{MFML}.
In practical terms this means that the $T$--matrix is calculated as in Eq.~(\ref{factor}),  
and only in the final expression $V$ is taken to depend 
on the energies $k^0$ and ${k'}^0$ of the external mesons.
This approximation can be justified by the observation that
any effect of the energy dependence of the vertices inside the loops can be 
absorbed in renormalization constants and does not influence the 
on--shell part of the $T$--matrix which is related to the physical 
amplitudes, Eq.~\ref{Tmatrixonshell}. 
For a more detailed discussion of these issues, see 
Refs.~\cite{Myhrer93,Myhrer94,ThorssonWirzba,WirzbaThorHirsch,MFML,OffShell1,OffShell2}. 

As a result of the above procedure, the problem reduces to the multiplication
and inversion of matrices spanned by the various meson--baryon channels.
Conceptually, this is of course much easier than solving the original 
Bethe--Salpeter equation.
Nevertheless, owing to the coupling of the various meson--baryon channels,
the expressions for the $T$--matrix elements in terms of the 
vertex-- and loop matrices $V$ and $J$  comprise several thousand terms each, 
and thus will not be given here explicitly.

%%--------------------------------------------------------------------------------

\subsection{Regularization}
\label{subsecRealpart}

We regularize the loop integral by employing twice subtracted dispersion
relations in the calculation of its real part. The subtraction parameters can
be fixed by fitting the resulting scattering amplitudes to the empirical
scattering lengths \cite{aKN_Iwasaki,NuclPhysB179198133}. 

This approach has some advantage over the use of cut--offs or form factors:
On the one hand, it keeps the vacuum--amplitude Lorentz--invariant.  
On the other hand it helps to overcome a technical problem of the 
selfconsistency procedure we aim at: Iteratively using the scattering matrix 
and the modified antikaon propagator as mutual input (see 
Fig.~\ref{overview_fig}) requires the knowledge of these functions on infinite
energy and momentum intervals. At large momenta, the antikaon propagator can be
approximated by the free one. However, it is not immediately clear how to 
treat the scattering amplitude in that respect. 
Use of a cut--off will introduce artificial peaks into 
the $T$--matrix, especially at higher energies and momenta. 
This is not the case for the amplitude obtained by a twice 
subtracted dispersion relation, which is smooth enough to be restricted to a 
finite interval as well.  

We find that our vacuum scattering amplitudes are in good agreement with the
$\Kbar N$ amplitudes of \cite{MFML}.

%%----------------------------------------------------------------------------------------

\subsection{Pion--nucleon interaction}
\label{pions}

The formulation of the $\Kbar N$ scattering problem in terms of 
coupled meson--baryon channels is the key ingredient of our model.
In particular, the $\Lambda(1405)$ will only arise as a dynamically generated 
resonance if the $\Kbar N$ channel is coupled to the $\pi \Sigma$ channel.
Consequently, the in--medium amplitude will feel any medium--modifications 
of the pion. 

The behavior of the pion in dense matter has been studied in great detail. 
The main sources of the pion selfenergy  are particle--hole and
delta ($\Delta(1232)$)--hole excitations. 
To capture these and other important features, 
in the pion sector we do not stick to the Lagrangian set up in 
Sect.~\ref{interaction} but include additional phenomenological terms 
which are taken from the literature. 
Here we basically follow Ref.~\cite{MUPionPaper}. 

The $p$--wave pion--nucleon interaction can be obtained from ${\cal L}_{pw}$  
in Eq.~(\ref{LMB1}). It is given by
\beq
{\cal L}_{\pi N}\;  =\; \frac{f_N}{{m_{\pi}}}\, 
                     \bar{\psi}\,\gamma^5\,\gamma^{\mu}\,\,
                   \vec{\tau}\,\psi\,\cdot\, \partial_{\mu}\,\vec{\phi}, 
\label{LPiN}
\eeq
with the pseudovector coupling constant $f_N = 1.01$.

The coupling to the $\Delta$ looks quite similar:
\beq
{\cal L}_{\pi N\Delta}=-\frac{f_{\Delta}}{{m_{\pi}}}
  \bar{\psi}\vec{T}^{\dag}\psi_{\mu}\cdot\partial^{\mu}\vec{\phi}
 \quad +\quad\mbox{h.~c.}
\label{LPiNDelta}
\eeq
Both, the $\pi NN$ and $\pi N\Delta$ vertices, are supplemented by a 
monopole form factor 
\beq
\Gamma_{\pi}(\vec{k}) = \frac{\Lambda^2}{\Lambda^2+\vec{k}^2}~,
\eeq
with $\Lambda = 550 \text{MeV}$. 
Finally,
short--range correlations are taken into account via Migdal parameters
$g'_{NN} = 0.8$ and $g'_{N\Delta} = g'_{\Delta\Delta} = 0.5$. No indications 
for $p$--wave pion condensation are observed for
the densities considered here (up to $5 \rho_0$).

Having calculated the pion propagator in dense matter, this propagator 
is put into the loops with the $\Lambda$ and $\Sigma$
baryons. These loops are then coupled to the other meson--baryon channels to
obtain the $T$--matrix element. As pointed out before,
we do not further modify the pion itself, i.e., we do not aim for 
selfconsistency in the case of the pion. 
This is justified because the above model of the pion
\cite{MUPionPaper,MURhoPaper,MUCorrelatorsPaper,MURhoA1Paper} is already 
designed to give a realistic result as it stands,  without further iteration.

%%===================================================================
  
\section{The kaon propagator in symmetric nuclear matter}
\label{Kaon_selfenergy}

We can now proceed to calculate the 
in--medium antikaon propagator following the selfconsistent scheme
outlined in the Fig.~\ref{overview_fig}.

\begin{figure}[htp]
\begin{center}
\hspace*{-4mm}\includegraphics[width=10.4cm,height=7cm]{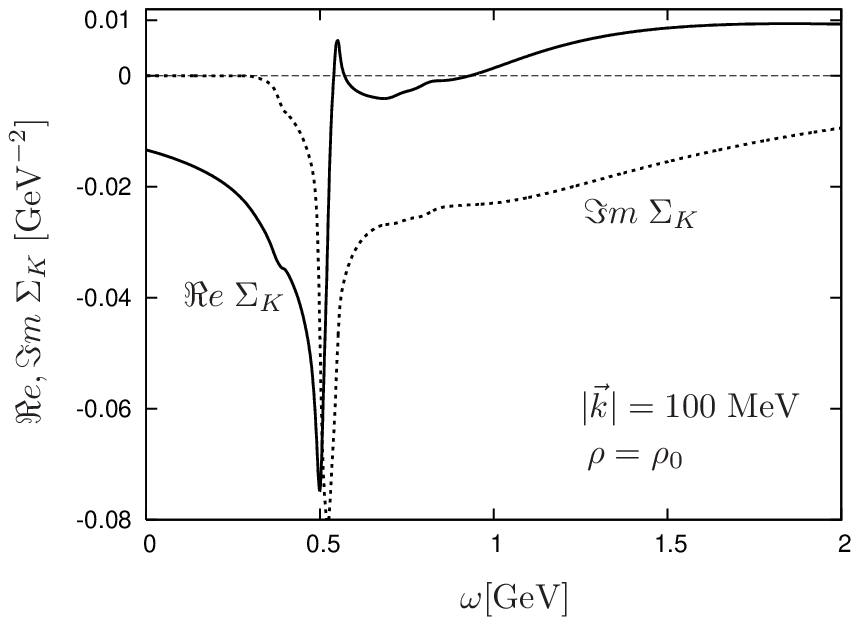}
\includegraphics[width=10cm]{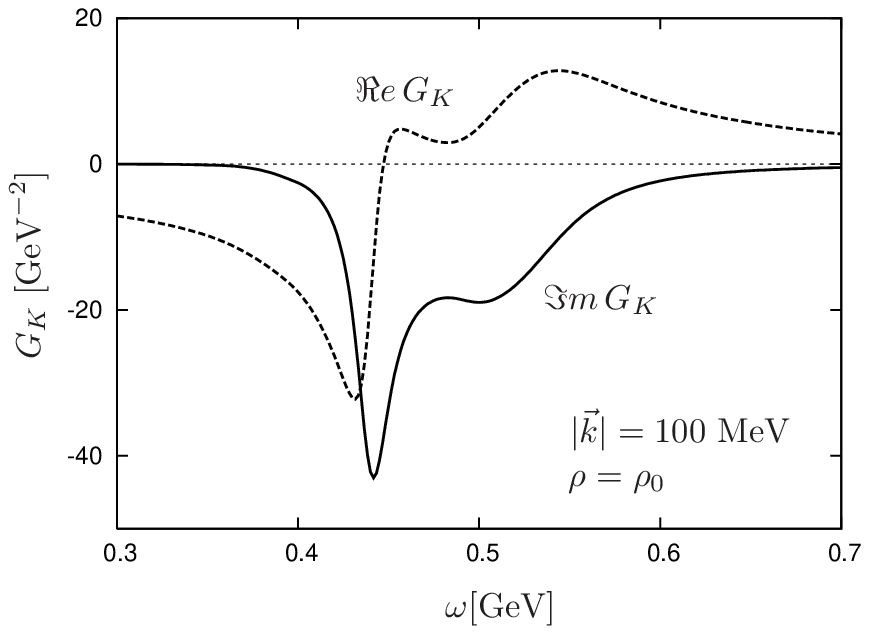}
\caption[Antikaon selfenergy: $\Sigma(\omega,\vec{k}=100$~MeV at $\rho = 
\rho_0$]{\small 
         Real part (solid) and imaginary part (dotted) of the antikaon   
         selfenergy (upper panel) and antikaon propagator (lower panel)
         as functions of energy $\omega$ at a three--momentum   
         of $|\vec{k}|=100$~MeV and a density of $\rho =  \rho_0$.}  
\label{plot_SigK_GK}
\end{center}
\end{figure}

A typical example for the resulting antikaon selfenergy is given in
the upper panel of Fig.~\ref{plot_SigK_GK}, 
where the real and imaginary parts of $\Sigma_K$
at normal nuclear matter density, $\rho = \rho_0 = 0.16$~fm$^{-3}$, are displayed as 
functions of energy at a finite three--momentum of $100$ MeV. 
The corresponding antikaon propagator $G_K$ is shown in the lower panel. 
The figure displays the typical properties of the antikaon when it is  
modified by the scattering processes in the medium.   

%%%%%%%%%%%%%%%%%%%%%%%%%%%%%%%%%%%%%%%%%%
%%%%%%%%%%%%%%%%%%%%%%%%%%%%%%%%%%%%%%%%%

The main peak of the spectral function (the imaginary part of $G_K$) is 
shifted to lower energies compared with its vacuum position at 
$\omega = \sqrt{m_K^2 + k^2} \simeq 505$~MeV.
However, this mass shift is not dramatic. A much
stronger effect is the broadening. Taking the full width at half maximum, 
the antikaon is ``spread" over roughly 130 MeV. 

In this section we want to investigate in some more details
which are the main physical sources for this behavior 
and how this depends on the kaon--momentum and on the density 
of the surrounding medium. 
We thereby concentrate on symmetric nuclear matter. Asymmetric matter
will be discussed in Sect.~\ref{chapAsymm}.

%%----------------------------------------------------------------------------------
\subsection{Effect of selfconsistency}
\label{sec_effectIteration}

In this subsection we briefly discuss the relevance of finding
a selfconsistent solution, rather than dressing the kaon 
perturbatively. 
Technically, once the selfconsistency calculation is set up and stable, 
the procedure works very well, converging after 3--4 iterations. 

\begin{figure}[ht]
\begin{center}
\includegraphics[width=10cm,height=7cm]{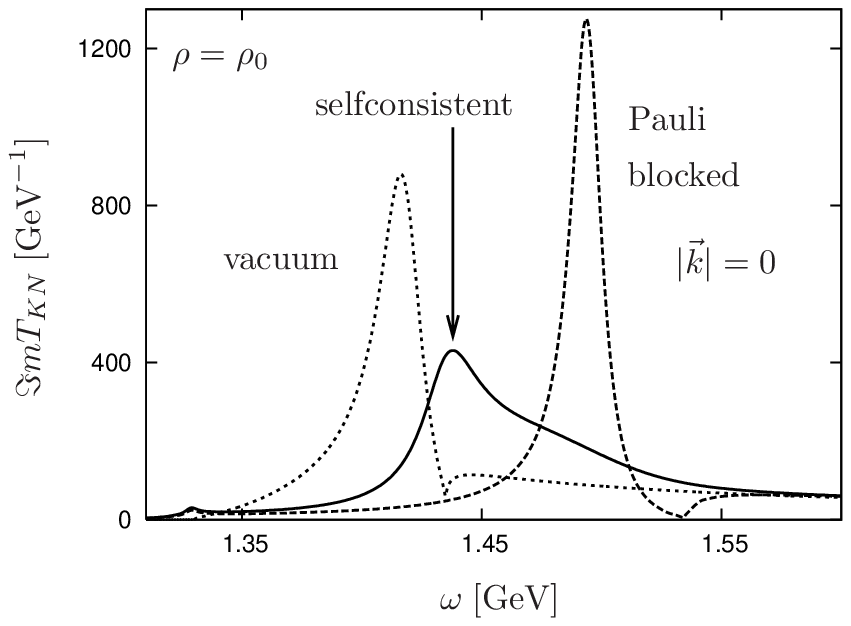}
\includegraphics[width=10cm,height=7cm]{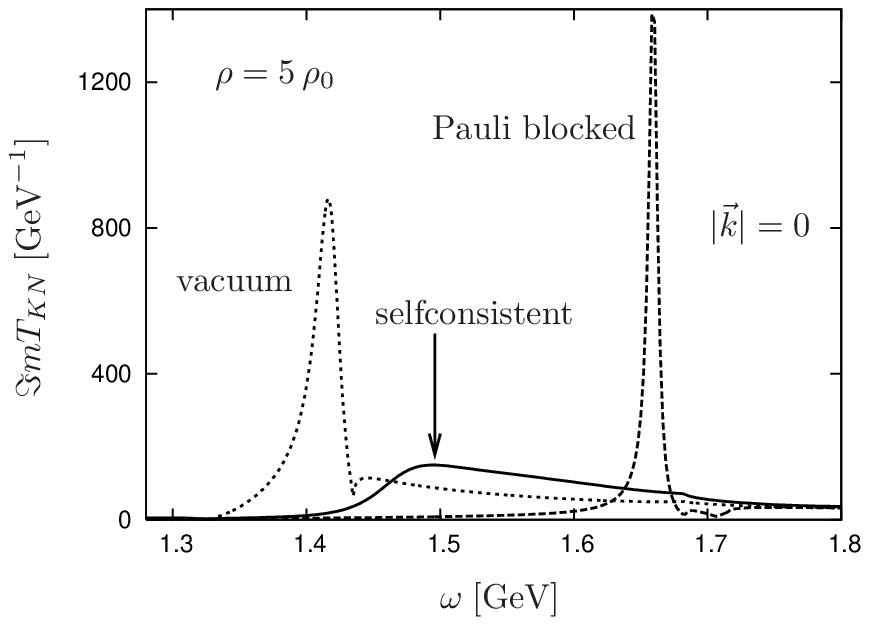}
\caption[$T$--matrix: ${\Im}m T$ before/after selfconsistency, $\rho=\rho_0$]
{\small
  Imaginary part of the $T$--matrix ${\Im}m T_{\Kbar N}$ at
  $|\vec{k}| = 0$ as a function of energy.
  Upper panel: $\rho= \rho_0$, lower panel: $\rho= 5\,\rho_0$.
  Dotted lines: vacuum result, dashed lines: in--medium
  result, only Pauli--blocking, solid lines: result after selfconsistency 
  is reached.}  
\label{GK5050k0selfcons}
\end{center}
\end{figure}

The effect of the selfconsistency iterations on the
$\Kbar N$ scattering amplitude is shown in Fig.~\ref{GK5050k0selfcons}
where the imaginary parts of $T_{KN}$ for vanishing three--momentum
and densities  $\rho = \rho_0$ (upper panel) and $5 \rho_0$ (lower panel) 
are plotted as a functions of the total energy. 
For comparison we also show the vacuum result (dotted lines). 
Here one can clearly see the $\Kbar N$ threshold at 1.43~GeV and the 
$\Lambda(1405)$ resonance below threshold. 
When the density effects are included by taking into account the 
Pauli blocking of the nucleon but using the unmodified kaon propagator
one arrives at the dashed lines. 
Because of the Pauli blocking the $\Kbar N$ threshold is shifted to higher
energies and, as a result, the $\Lambda(1405)$ is also pushed 
upwards.

However, in the selfconsistent calculation (solid lines in Fig.~\ref{GK5050k0selfcons}),
the change in the antikaon mass and especially the 
broadening of the kaon spectral function (see below)
re--enables scattering into lower 
lying states and the resonance is pulled down again.
This behavior has been discussed first in Ref.~\cite{Lutz,LutzKorpa}.

%%%%%%%%%%%%%%%%%%%%%%%%%%%%%%%%%%%%%%%%
%%%%%%%%%%%%%%%%%%%%%%%%%%%%%%%%%%%%%%%%%%

\begin{figure}[htp]
 \begin{center}
\includegraphics[width=10cm]{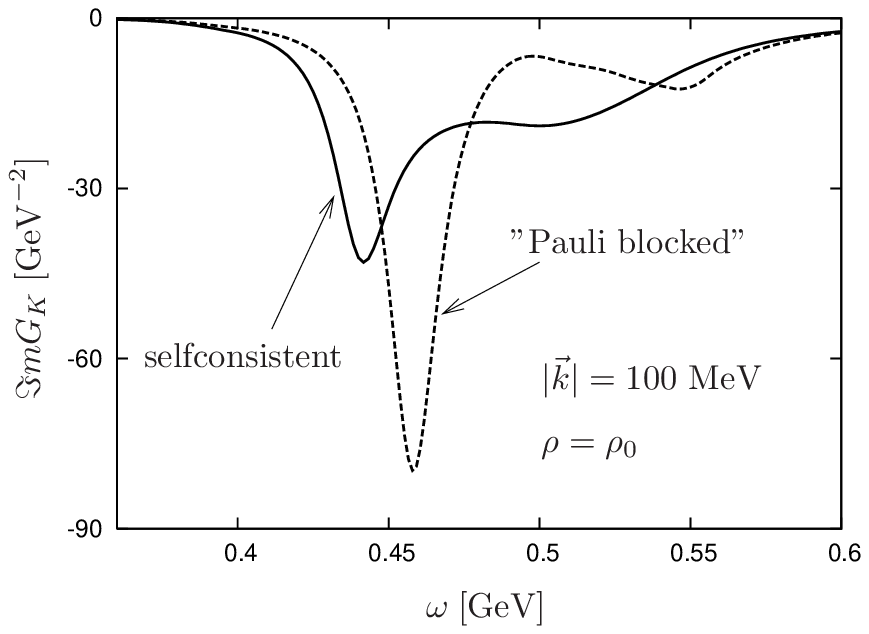}
\includegraphics[width=10cm]{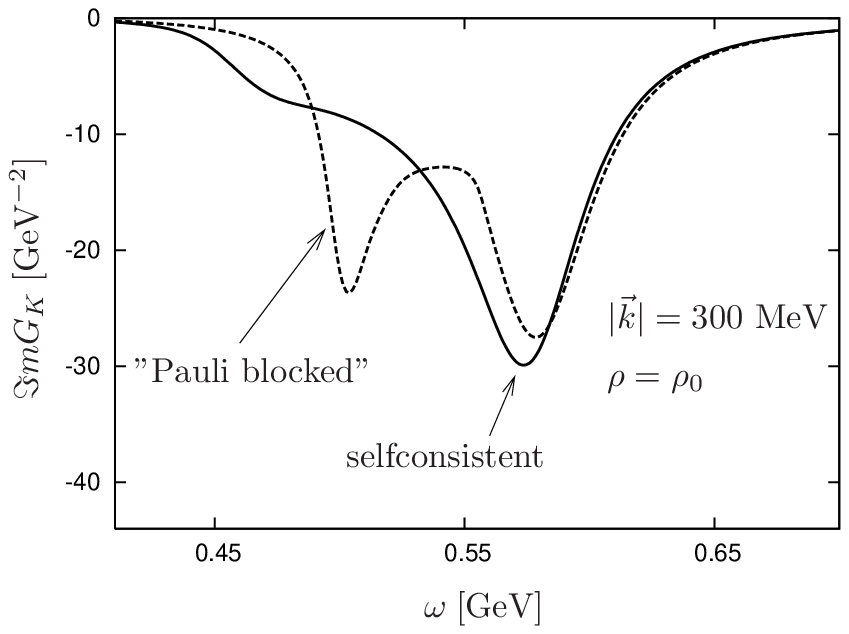}
\caption[Antikaon propagator: ${\Im}m G_{K}$ before/after selfconsistency, 
$|\vec{k}| = 100$ MeV]
{\small Imaginary part of the antikaon propagator ${\Im}m G_{K}$ at normal 
  nuclear density $\rho=\rho_0$ as functions of energy.
  Upper panel: $|\vec{k}| = 100$ MeV, lower panel: $|\vec{k}| = 300$ MeV.
  Dashed lines: after one iteration, solid lines: after  selfconsistency is 
  reached.}
\label{GK5050k100selfcons}
 \end{center}
\end{figure}

Fig.~\ref{GK5050k100selfcons} shows the 
corresponding effect of the selfconsistency iteration on
the antikaon propagator (imaginary part) at $\rho = \rho_0$
and two different 3--momenta.
Again, the dashed lines, labeled ``Pauli--blocked", mark the results after 
the first iteration step, i.e., when only the Pauli--blocked $T$--matrix 
elements (cf. dashed lines in Fig.~\ref{GK5050k0selfcons}) 
are used in the selfenergy integrals. 
The selfconsistency iterations broaden the antikaon and shift it further down
(solid line).

Also note that in the ``Pauli--blocked" calculation 
there is a double peak structure in ${\Im}m G_{\Kbar}$ which 
becomes more prominent at higher momentum.
This feature is largely washed out by the selfconsistency iterations.

%%%%%%%%%%%%%%%%%%%%%%%%%%%%%%%%%%%%%%%%
%%%%%%%%%%%%%%%%%%%%%%%%%%%%%%%%%%%%%%%%

%------------------------------------------------------------------------------
\subsection{Momentum and density dependence of the antikaon propagator}
\label{subsecMomentumdependence}

\begin{figure}[htp]
\begin{center}
\includegraphics[width=10cm]{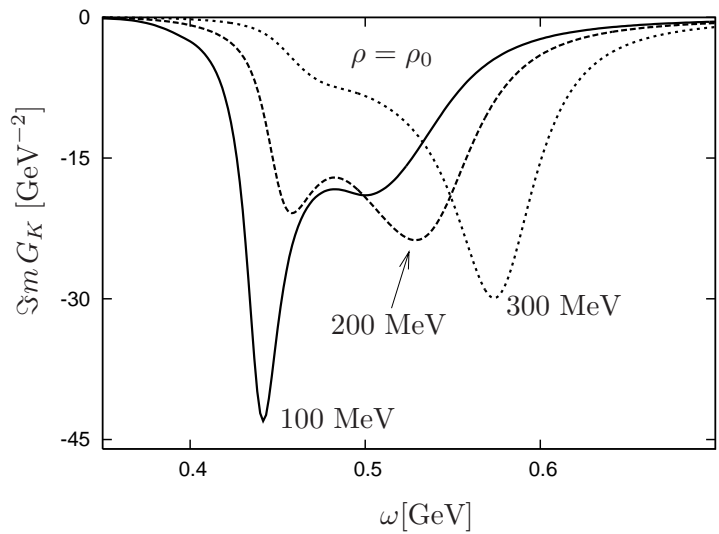}
\caption[Pion propagator $G_\pi$]
{\small Imaginary part of the selfconsistent antikaon propagator at  density $\rho = \rho_0$
        as function of energy. The various curves correspond to three
        different 3--momenta: $|\vec{k}| = 100$~MeV (solid), 
        200~MeV (dashed), and 300~MeV (dotted).
} 
\label{CompareGK_5050_low_k}
\end{center}
\end{figure}

Fig.~\ref{GK5050k100selfcons} already indicates 
the structure of the antikaon spectral function as a function of momentum. 
This is summarized in Fig.~\ref{CompareGK_5050_low_k} which shows the imaginary part of 
the kaon propagator at density $\rho = \rho_0$ for three different 
3--momenta, $|\vec{k}| =$ 100~MeV, 200~MeV, and 300 MeV.  

As mentioned above, one typically finds two peaks.
It turns out that for low values of $|\vec{k}|$ most of the strength is 
contained in the lower peak (solid line). With increasing $|\vec{k}|$,
this peak becomes gradually reduced while the upper peak grows (dashed line). 
Finally, at large momenta, most of the strength is found in the upper peak
(dotted line). 

%%%%%%%%%%%%%%%%%%%%%%%%%%%%%%%%%%%%%%%%%%%%
%%%%%%%%%%%%%%%%%%%%%%%%%%%%%%%%%%%%%%%%%%%%

The nature of the two--peak structure can be easily understood in 
terms of a simple two--level model, describing the coupling of a 
bare antikaon state to a $\Lambda(1405)$--hole state. 
At low momenta the unperturbed energies of these states are very
close to each other, the $\Lambda(1405)$--hole state being only about
30~MeV below the bare kaon. However, turning on the interaction,
the splitting of the peaks becomes larger through 
level repulsion, which shifts the ``antikaon peak'' somewhat up and the
``$\Lambda(1405)$--hole peak'' further down in energy.
The latter receives most of the strength at low momenta
while at large momenta the kaon branch dominates. 
This is quite analogous to the well--known case of a pion coupled
to a $\Delta$--hole state \cite{MUPionPaper}.
However, it should be noted that this picture is rather crude
since, as a consequence of our coupled--channel approach, we cannot
resolve the peaks in the propagator 
into particular quasiparticle states. 

\begin{figure}[htp]
\begin{center}
\includegraphics[width=10cm]{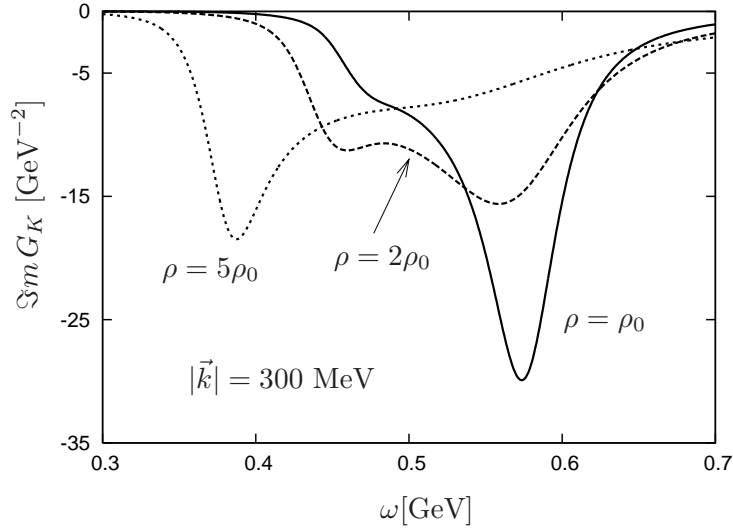}
\caption[Branches of ${\Im}m G_{{\Kbar}}$, $|\vec{k}|=300$ MeV]
{\small Imaginary part of the selfconsistent antikaon propagator at 3--momentum 
        $|\vec{k}| = 300$~MeV. The various curves correspond to three
        different densities: $\rho = \rho_0$ (solid), 
        $2\rho_0$ (dashed), and $5\rho_0$ (dotted).
}
\label{CompareGK_u_5050_iteration_k030}
\end{center}
\end{figure}

The above features of the kaon propagator are subject to a strong variation 
with density. This is illustrated in Fig.~\ref{CompareGK_u_5050_iteration_k030} where 
the imaginary part of the kaon propagator is displayed as a function of
energy at fixed 3--momentum $|\vec{k}|= 300$ MeV and
the three densities $\rho = \rho_0, \,2\,\rho_0$, and $5\,\rho_0$.
Obviously, the strength that at $\rho = \rho_0$ is found in the 
higher--lying peak is still in the lower branch at the higher densities. 
This behavior is consistent: At higher densities the medium modifications 
are stronger. Thus the effect of increasing 3--momentum as discussed above 
sets in only at even higher momenta.

%%%%%%%%%%%%%%%%%%%%%%%%%%%%%%%%%%%%%%%
%%%%%%%%%%%%%%%%%%%%%%%%%%%%%%%%%%%%%%%%%%%

At very high momenta,  the free dispersion relation of the antikaon is 
regained. This provides some technical help, allowing the use of the free 
antikaon beyond $\omega = 1.5$ GeV and $|\vec{k}| = 1.2$ GeV.

%%--------------------------------------------------------------------------------

\subsection{In--medium modification of the pions}
\label{chapPions}

The importance of the pion in the $\Kbar N$ scattering processes was already 
emphasized. 
As pointed out earlier, the pion undergoes considerable 
modifications when placed in dense matter.
In the following, we want to study the effect of these medium modifications
of the pion propagator on the kaon propagator through the coupled--channel
mechanism.  

\begin{figure}[htp]
\begin{center}
\includegraphics[width=8.5cm]{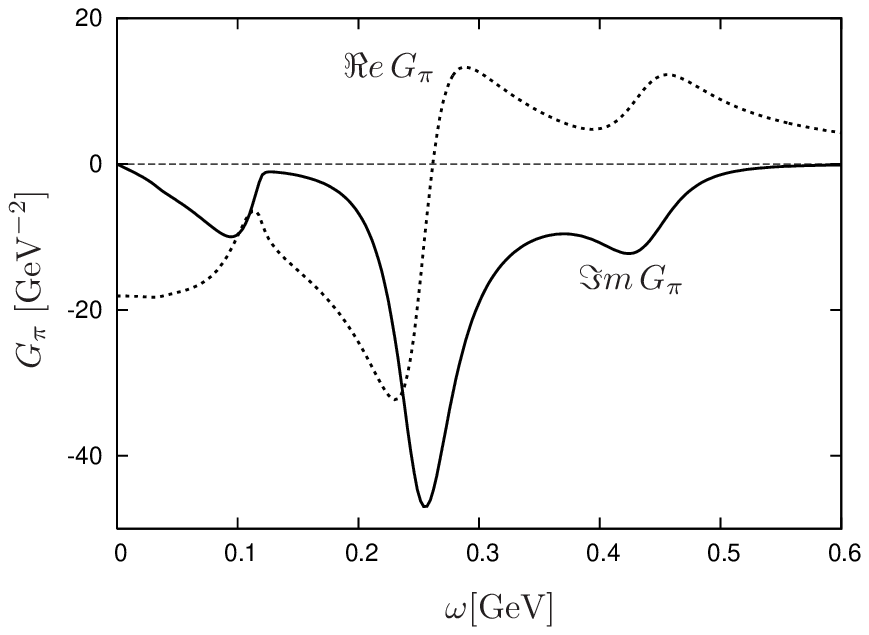}
\hspace*{-3mm}\includegraphics[width=8.8cm]{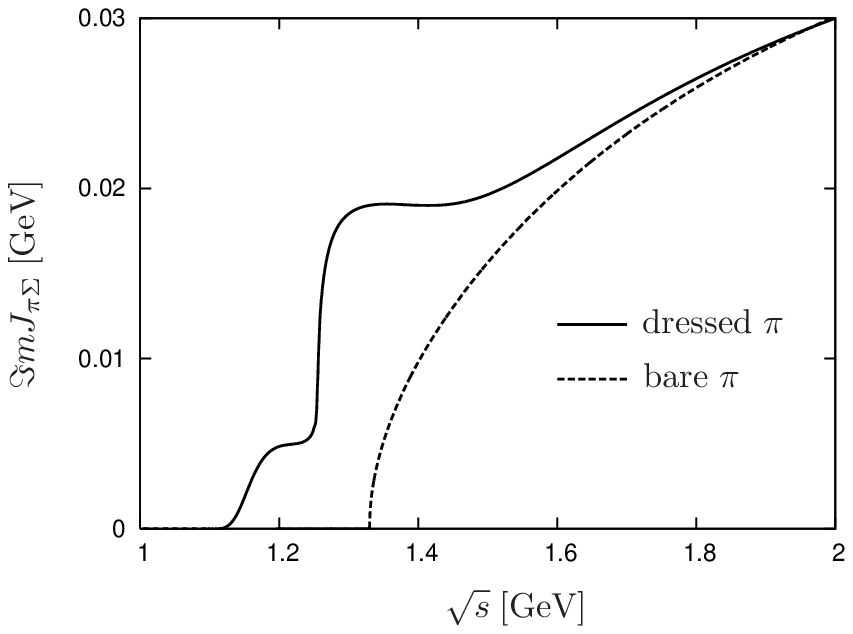}
\includegraphics[width=8.5cm]{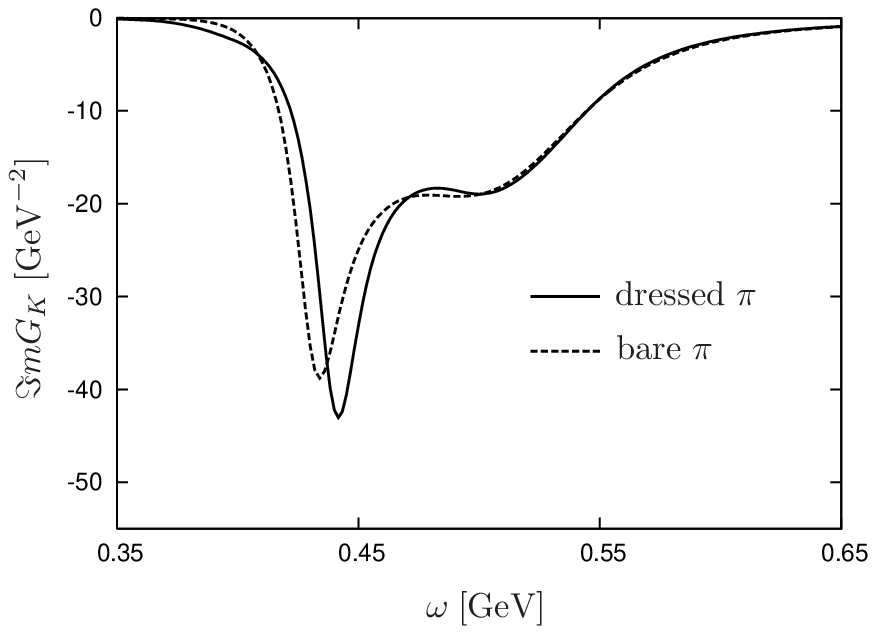}
\caption[Pion propagator $G_\pi$]
{\small 
Upper panel:
Real part (dotted line) and imaginary part (solid line) of the 
pion propagator at $|\vec{k}| = 300$~MeV as functions of energy.\\
Central panel:
Imaginary part of the $\pi \Sigma$ loop function
at $|\vec{q}|=100$~MeV
as function of the invariant mass, incorporating dressed 
(dashed line) and bare (solid line) pion propagators.\\
Lower panel:
Imaginary part of the antikaon propagator
 at $|\vec k| = 100$~MeV
as function of energy,
incorporating dressed (solid line) and bare (dashed line) pions.\\
All calculations in this figure have been performed for $\rho = \rho_0$.
} 
\label{Gpi_k030a}
\end{center}
\end{figure}

Our results are summarized in Fig.~\ref{Gpi_k030a}.
The pion propagator at normal nuclear matter density is displayed in
the upper panel. At the indicated 
pion 3--momentum of $300$ MeV, the broadening effect is maximal.

%%%%%%%%%%%%%%%%%%%%%%%%%%%%%%%%%%%%%%%%%%%
%%%%%%%%%%%%%%%%%%%%%%%%%%%%%%%%%%%%%%%%%%

Typically, the broadening of the pion in dense matter has some bearing 
on the $\pi$--baryon loop function in the form of a bump around threshold. 
As an example, the central panel of Fig.~\ref{Gpi_k030a}
shows the imaginary part of 
the $\pi \Sigma$ channel -- loop function, ${\Im}m J_{\pi \Sigma}$, 
at normal nuclear matter density and three--momentum $|\vec q| = 100$~MeV.
The effect seems to be sizeable, the difference to the vacuum curve is 
quite pronounced. 

On the other hand, the lower panel indicates the difference 
it makes for the imaginary part of the kaon propagator, ${\Im}m G_{\Kbar}$, 
whether the bare or the dressed pion propagator is used as input. 
The figure shows the selfconsistent result, again at normal nuclear matter density and $|\vec k| = 100$~MeV.
Obviously, very little is left of the effect of pion dressing, in agreement
with \cite{LutzKorpa0404088}, while in \cite{Tolos0007042} a stronger
effect has been found.
The reason is the strong momentum dependence of the pion selfenergy. 
At the $|\vec k| = 300$ MeV as in the upper panel, its effect is most 
pronounced. However, already the loop function (central panel) 
is obtained by integrating  the in--medium pion  over all momenta. 
The kaon propagator (lower panel) is the result of yet another 
integration -- not counting the various iterations to achieve a
selfconsistent result.

%%--------------------------------------------------------------------------------------------------

\subsection{$p$--wave antikaon selfenergy}
\label{pwavekaonselfenergy}

As mentioned earlier, we neglect 
contributions to the antikaon selfenergy arising from the $p$--wave
vertices of Eq.~\eqref{LMB1} in our model.
The reason is that these contributions are expected to be small, 
but would considerably complicate the selfconsistency problem.

\begin{figure}[htp]
\begin{center}
  \includegraphics[width=6cm]{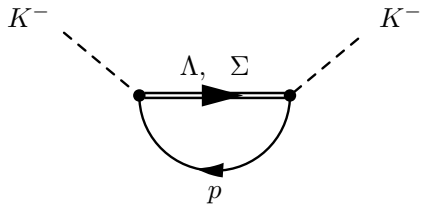}
  \caption[$p$--wave selfenergy]
   {\small Diagrams corresponding to the 
   $p$--wave contribution to the antikaon selfenergy.}
   \label{SigK_pwave}
  \end{center}
\end{figure}

To check the (un--) importance of the $p$--wave vertices for the kaon propagator, we include their 
 contributions to the selfenergy, but only in a perturbative way.
This means, we first calculate the selfconsistent kaon propagator obtained
with $s$--wave interactions only, as before.  
Then we add the $p$--wave contributions to the selfenergy, i.e., 
$\Lambda$--hole and $\Sigma$--hole diagrams as depicted in 
Fig.~\ref{SigK_pwave}. As in the pion case we use a monopole form factor with $\Lambda = 550$ MeV.
For the Migdal parameters we take the classical value of $g' = 1/3$.
The resulting total selfenergy  is then used
to calculate the modified kaon propagator, but the procedure is not 
iterated again.

%%%%%%%%%%%%%%%%%%%%%%%%%%%%%%%%%%%%%%%%%%%%%%%
%%%%%%%%%%%%%%%%%%%%%%%%%%%%%%%%%%%%%%%%%%%%%

%%%%%%%%%%%%%%%%%%%%%%%%%%%%%%%%%%%%%%%
\begin{figure}[htp]
\begin{center}
  \includegraphics[width=10cm]{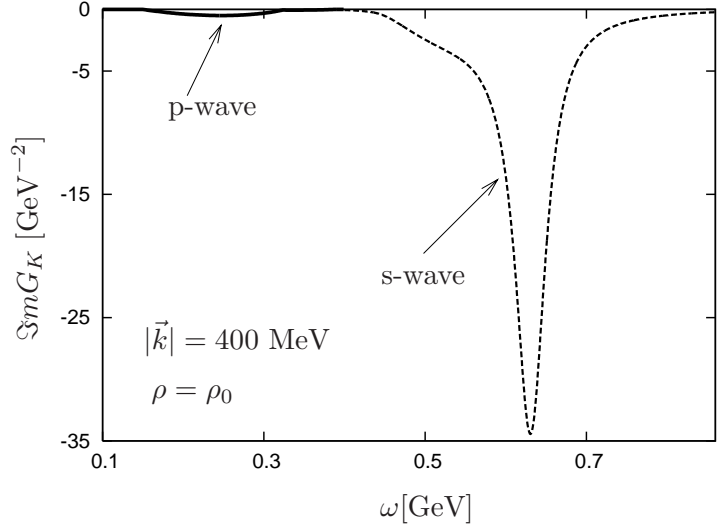}
   \caption[$K^-$ $s+p$--wave propagator]
       {\small Imaginary part of
       the $K^-$  propagator at $|\vec{k}|=400$~MeV and $\rho =  \rho_0$:
       $s$--wave (dashed) and $p$--wave (solid) contributions.}
   \label{ImGK_pwave}
  \end{center}
\end{figure}
%%%%%%%%%%%%%%%%%%%%%%%%%%%%%%%%%%%%%%%%%%%%%

The result for $\rho = \rho_0$ and $|\vec{k}|=400$~MeV is shown in 
Fig.~\ref{ImGK_pwave}. 
Although the chosen three--momentum roughly corresponds to the value
where the $p$--wave effect has its maximum, it is obviously very small.
This is due to the position of the $p$--wave selfenergy at very low 
energies. In fact, it lies entirely in the space--like region. Thus the kaon
with its bare pole at $\omega \sim 600$ MeV does not feel much of the 
$p$--wave selfenergy. 
Hence, even though somewhat larger effects have been found when the 
$p$--wave interactions are included in the meson--baryon scattering 
equation \cite{Tolos0007042,LutzKorpa,KolVos}, their negligence seems to be justified.

%%==============================================================

\section{Asymmetric nuclear matter}
\label{chapAsymm}

In this section we extend our discussion to isospin asymmetric matter,
i.e., to the case of unequal densities of protons and neutrons.
This case is relevant for the description of neutron star matter which 
is expected to consist of roughly 90\% neutrons and 10\% protons
(see Sect.~\ref{chapKcondens}).
The calculation of the $\Kbar N$ scattering and the antikaon selfenergy in the
case of asymmetric nuclear matter follows the scheme developed in the previous
sections. To discuss the systematic behavior we have investigated
various proton--to--neutron ratios at various densities. 
In this context a somewhat more precise terminology is required: 
The density $\rho$ referred to is always the total baryon number density. 
Since we consider the zero--temperature case, $T=0$, the only baryons
occurring in sizeable number are the nucleons.\footnote{For a nucleon gas with $90 \%$ neutrons, 
the neutron Fermi energy reaches the mass of the \mbox{(undressed) $\Lambda$} only 
at $\rho=6.7 \rho_0$} Of these, a fraction $x_p$ consists of protons.
The symmetric case is thus characterized by $x_p = 0.5$.

%%--------------------------------------------------------------------------------------------------------

\subsection{Kaons in asymmetric matter}
\label{subsecKaonsinasym}

%%%%%%%%%%%%%%%%%%%%%%%%%%%%%%%%%%%%%%%%%%%
\begin{figure}[htp]
\begin{center}
\includegraphics[width=9.5cm]{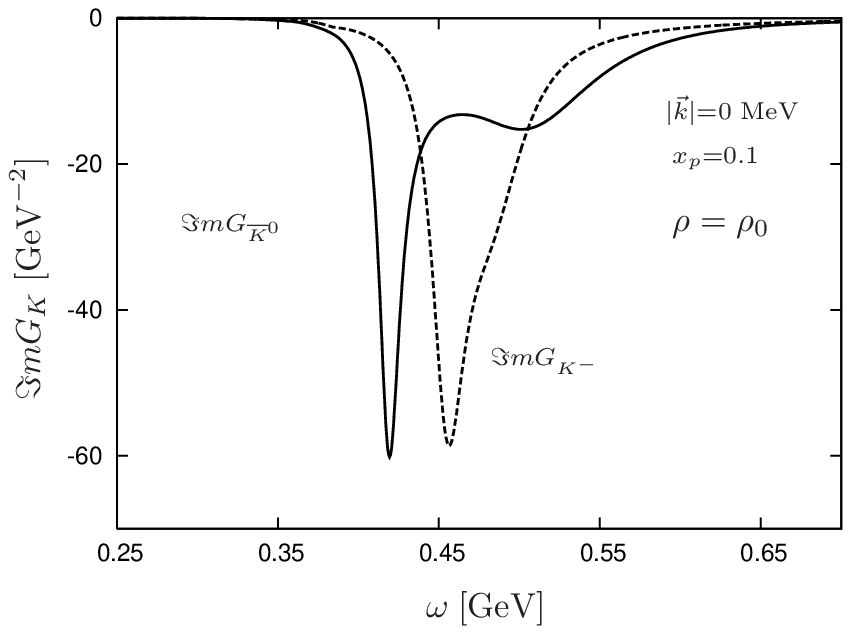}
\includegraphics[width=9.5cm]{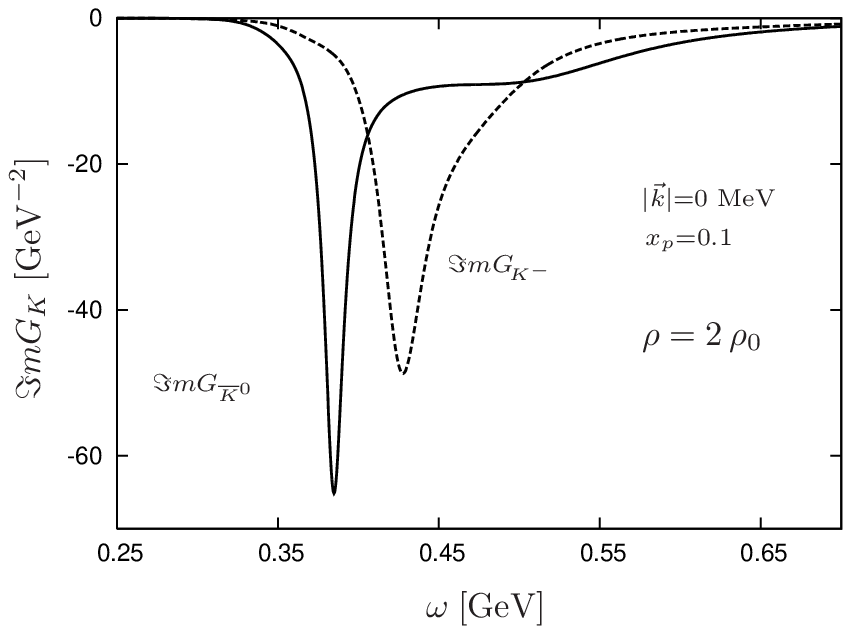}
\includegraphics[width=9.5cm]{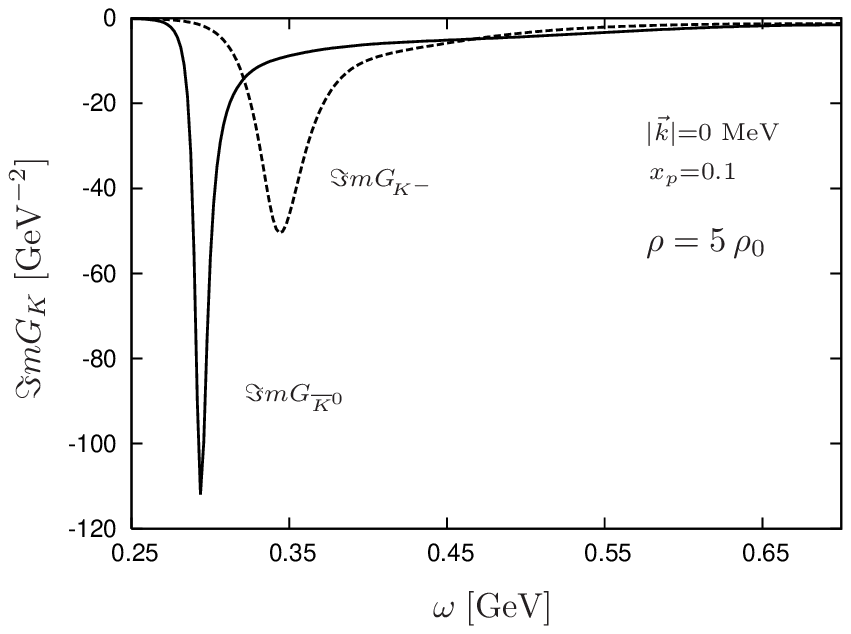}

\caption[Kaon propagator: ${\Im} m G_{\KObar}, {\Im} m G_{K^-}\;\; (u=1, u=2)$]
{\small
Imaginary parts of the $K^-$ (solid line ) and $\KObar$ (dashed line)
propagators as functions of energy for $|\vec{k}| = 0$ and
a proton fraction $x_p = 0.1$. Upper panel: $\rho = \rho_0$, 
central panel: $\rho  = 2\, \rho_0$, lower panel: $\rho = 5\rho_0$.}
\label{ImGKm0_u_10.90}
\end{center}
\end{figure}
%%%%%%%%%%%%%%%%%%%%%%%%%%%%%%%%%%%%%%%%%%%%%

In Fig.~\ref{ImGKm0_u_10.90}, the kaon spectral function at vanishing 
3--momentum is shown for a proton fraction of $10$\% and three 
different densities.
The most important difference to the symmetric case is the fact
that, due to the isospin breaking surrounding,
the propagators of $K^-$ and  $\KObar$ are no longer identical, but
the $\KObar$ is much more affected  by the dense medium. 
This can be explained by the fact that the Weinberg--Tomozawa term 
is strongest in the $\KObar n$ and $K^- p$ channels. Therefore, since
the neutron density is ten times higher than the proton density,
the $\KObar$ undergoes a stronger modification,
although the various contributions get thoroughly mixed in the 
selfconsistency iterations.

\begin{figure}[htp]
\begin{minipage}[h]{8.5cm}
\begin{center}
\includegraphics[width=8.5cm]{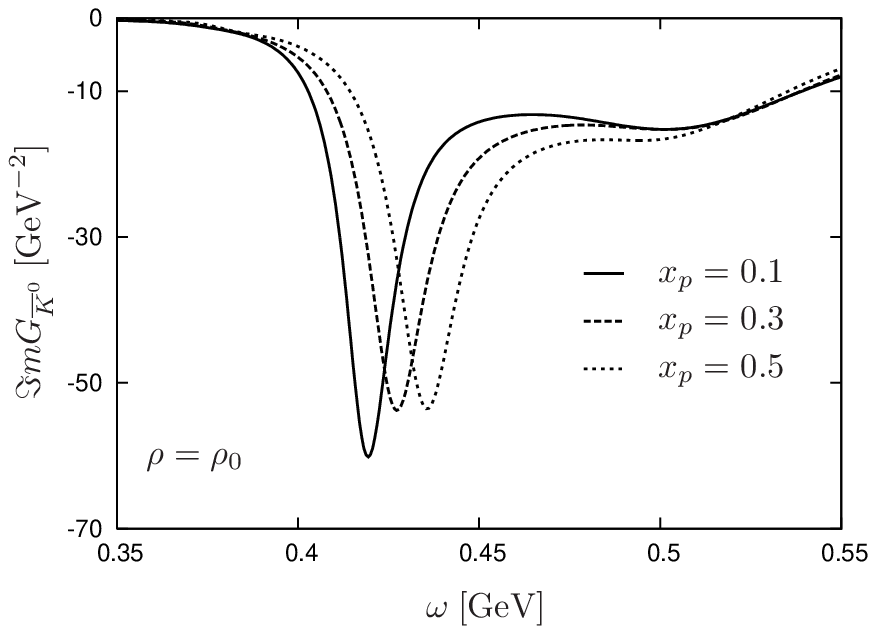}
\end{center}
\end{minipage}
\begin{minipage}[h]{8.5cm}
\begin{center}
\includegraphics[width=8.5cm]{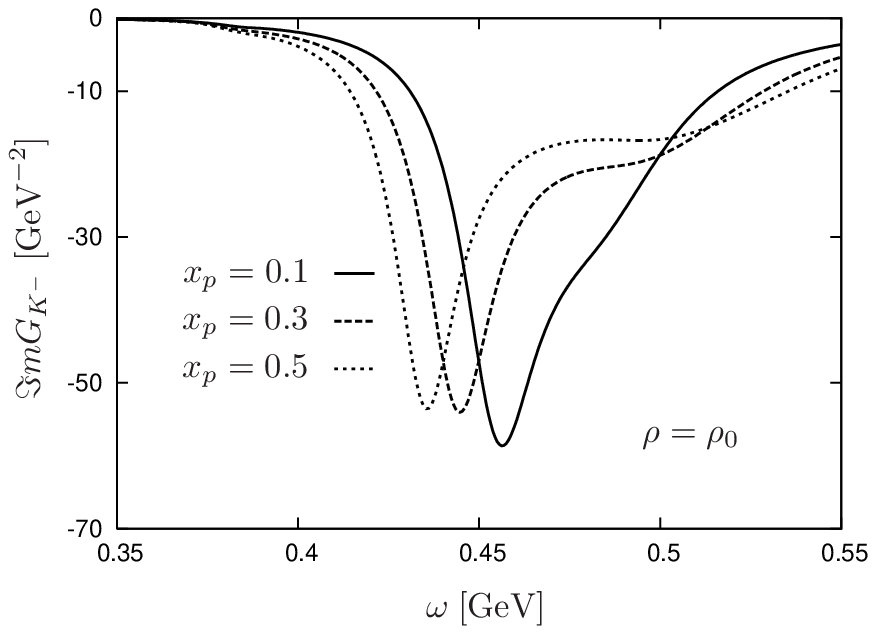}
\end{center}
\end{minipage}
%%%
\begin{minipage}[h]{8.5cm}
\begin{center}
\includegraphics[width=8.5cm]{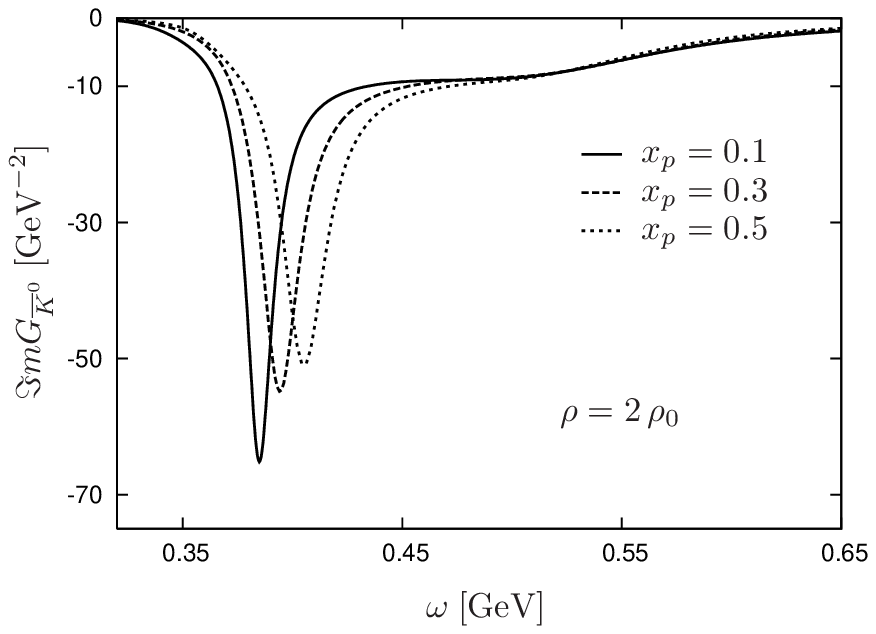}
\end{center}
\end{minipage}
\begin{minipage}[h]{8.5cm}
\begin{center}
\includegraphics[width=8.5cm]{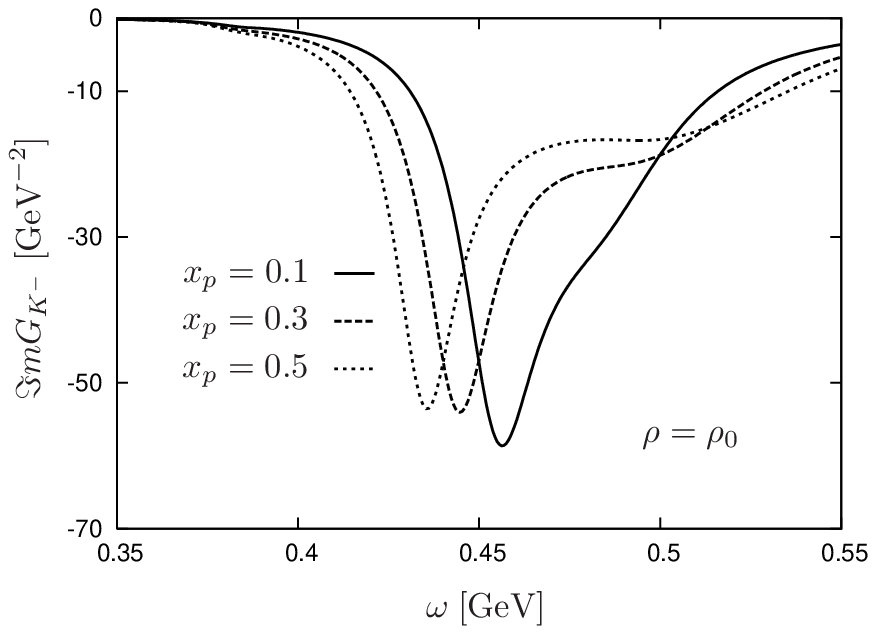}
\end{center}
\end{minipage}
%%%
\begin{minipage}[h]{8.5cm}
\begin{center}
\includegraphics[width=8.5cm]{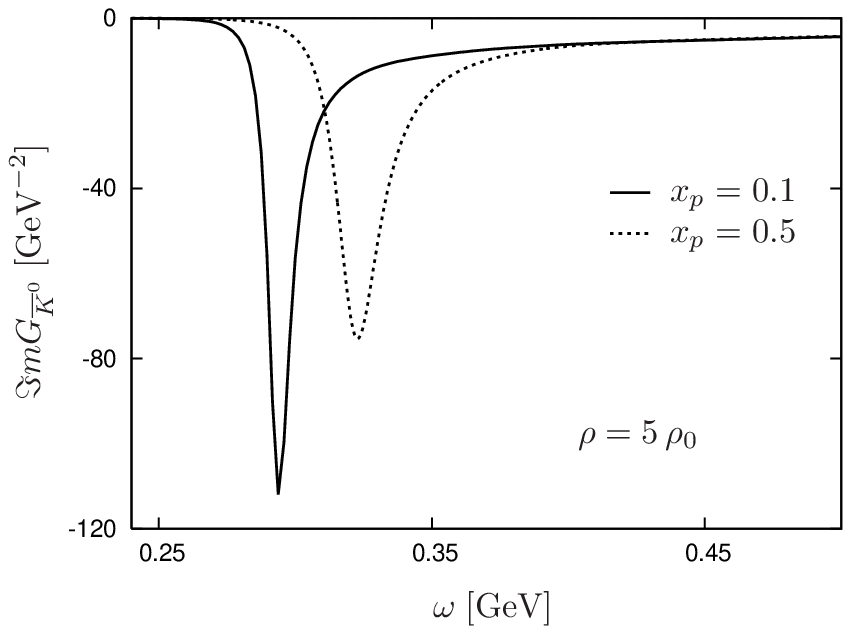}
\end{center}
\end{minipage}
\begin{minipage}[h]{8.5cm}
\begin{center}
\includegraphics[width=8.5cm]{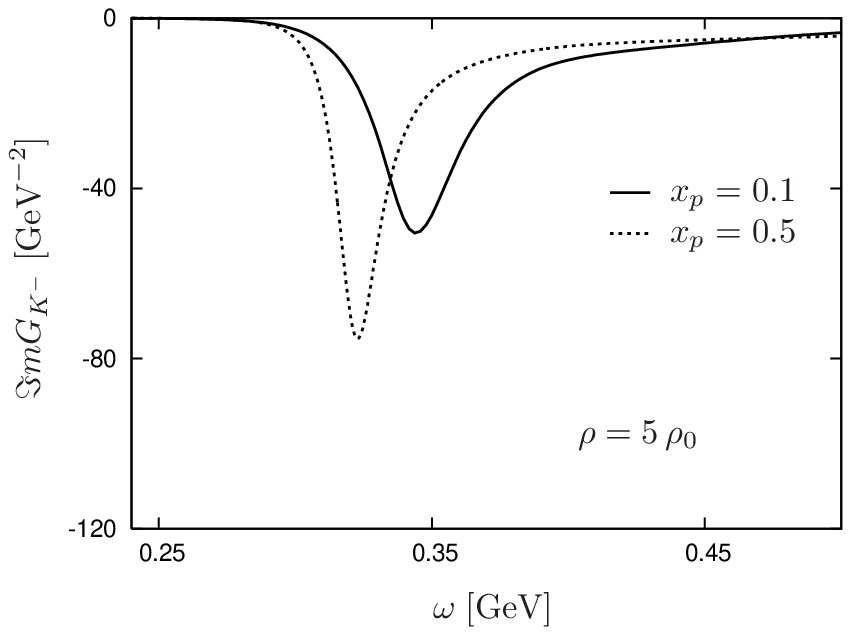}
\end{center}
\end{minipage}
%%%
 \caption[Kaon propagator: ${\Im} m G_{\KObar}, {\Im} m G_{K^-}\;\; (u=1)$]
{\small Imaginary parts of the propagators of $\KObar$ (left panels) 
and $K^-$ (right panels) as functions of energy at $|\vec k| = 0$
for different baryon densities and different proton fractions.}
 \label{ImGKm0_u1_xp}
\end{figure}

The variation of ${\Im} m G_K$ with the proton--neutron asymmetry 
at different densities can be
studied by comparing the curves in Fig.~\ref{ImGKm0_u1_xp}. 
This figure shows a smooth variation of the properties of the antikaons
with density and isospin composition.

%%%%%%%%%%%%%%%%%%%%%%%%%%%%%%%%%%%%%%%
%%%%%%%%%%%%%%%%%%%%%%%%%%%%%%%%%%%%%%%%%%%%%

\begin{figure}[htp] 
\begin{minipage}[h]{8.5cm}
\begin{center}
\includegraphics[width=8.5cm]{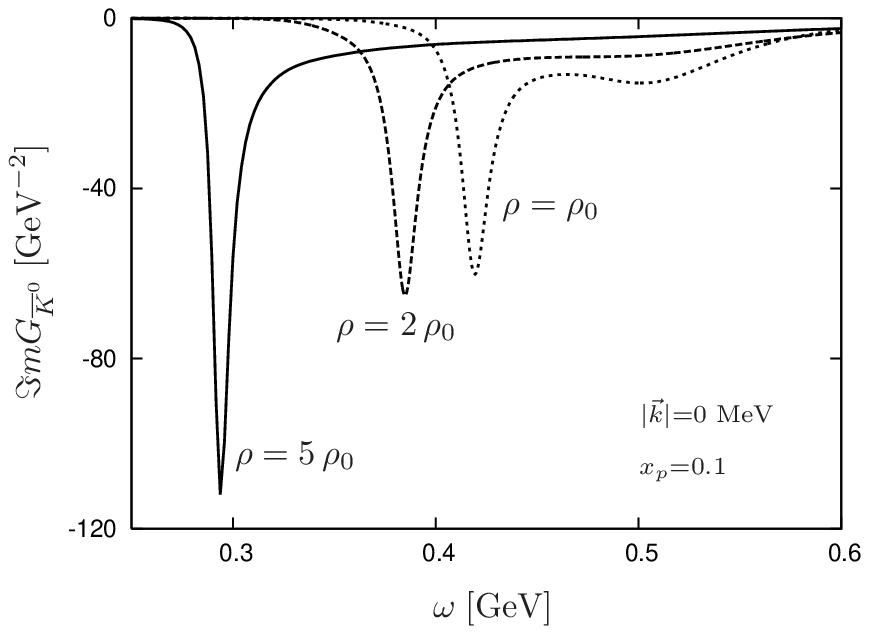}
\end{center}
\end{minipage}
\begin{minipage}[h]{8.5cm}
\begin{center}
\includegraphics[width=8.5cm,height=6.4cm]{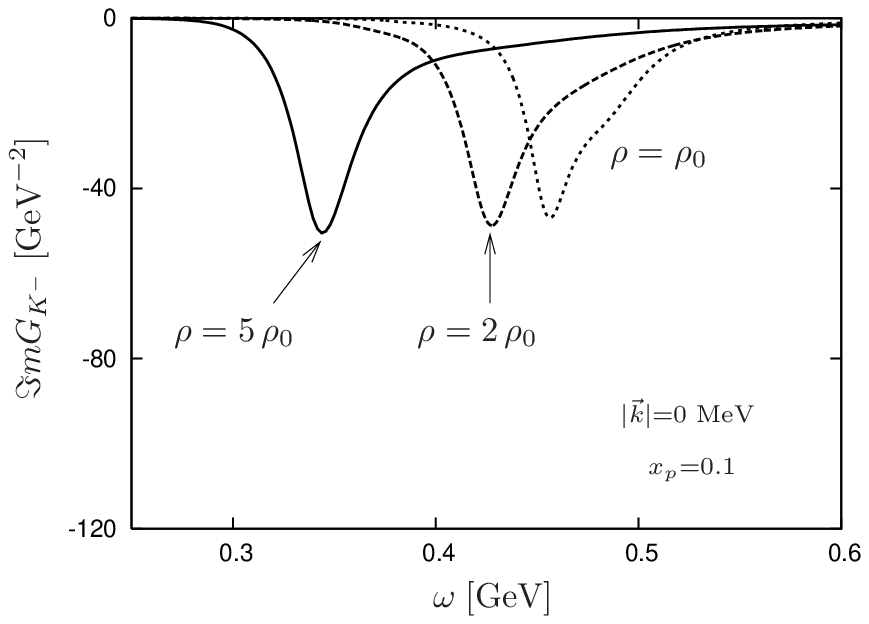}
\end{center}
\end{minipage}
 \caption[Kaon propagator: ${\Im} m G_{\KObar}, {\Im} m G_{K^-}\;\; (u=1,2,5)$]
{\small Imaginary parts of the antikaon propagators as functions of energy at 
  $x_p = 0.1, |\vec{k}| = 0$ for different
  densities. $\KObar$ on the left,  $K^-$ on the right.}
 \label{ImGKm0_u_10.90_k0}
\end{figure}

Finally, Fig.~\ref{ImGKm0_u_10.90_k0} shows the
imaginary part of the antikaon propagator, now comparing directly the different
densities, at an asymmetry of $x_p = 0.1$. As mentioned, the growing density affects the $\KObar$ 
much more than the $K^-$. However, the overall effect is not dramatic, the peaks of the 
spectral functions being shifted by no more than $\sim 200$ MeV from the vacuum 
positions at $500$ MeV. It also becomes obvious that the width of the antikaon shrinks 
again at high density. This is of course an effect of shrinking phasespace.

%%%%%%%%%%%%%%%%%%%%%%%%%%%%%%%%%%%%%%
%%%%%%%%%%%%%%%%%%%%%%%%%%%%%%%%%%%%%%%%%%%

%-----------------------------------------------------------------------------------------------------
\subsection{Effect of s--wave pion--nucleon interactions}
\label{subsecPionsAsym}

In Sect.~\ref{chapPions}
we have seen that the medium modifications of pions have a rather
small impact on the kaon propagator in symmetric nuclear matter.
We have thereby restricted ourselves to include p--wave 
pion--nucleon interactions, as outlined in Sect.~\ref{pions}.
However, there are also known s--wave interactions 
which we have neglected so far. 
The contribution of the Weinberg--Tomozawa term to the $s$--wave pion selfenergy is
usually held responsible for the existence of deeply bound pionic states in
heavy nuclei \cite{Gilg}. The complete expressions for asymmetric matter in
two--loop chiral perturbation theory are given by a calculation by Kaiser and
Weise \cite{pion_s_wave-Weise}. Their analytic expressions
depend mainly on the difference of neutron and proton Fermi momenta,
which means that the s--wave selfenergy is especially important in isospin 
asymmetric matter. 

\begin{figure}[htp]
  \begin{minipage}[h]{8.5cm}
  \begin{center}
  \includegraphics[width=8.5cm]{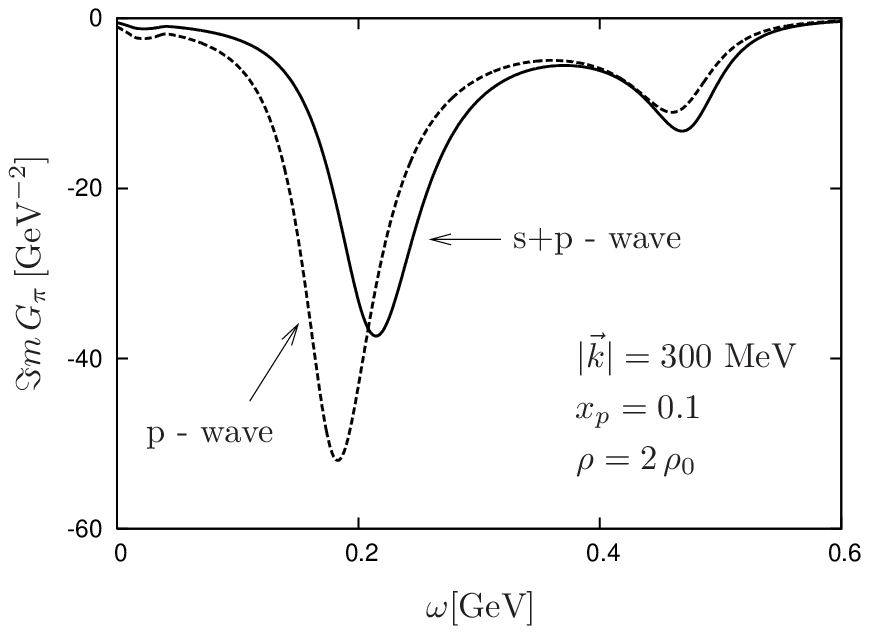}%}
  \end{center}
  \end{minipage}
  \begin{minipage}[h]{8.5cm}
  \begin{center}
  \includegraphics[width=8.5cm]{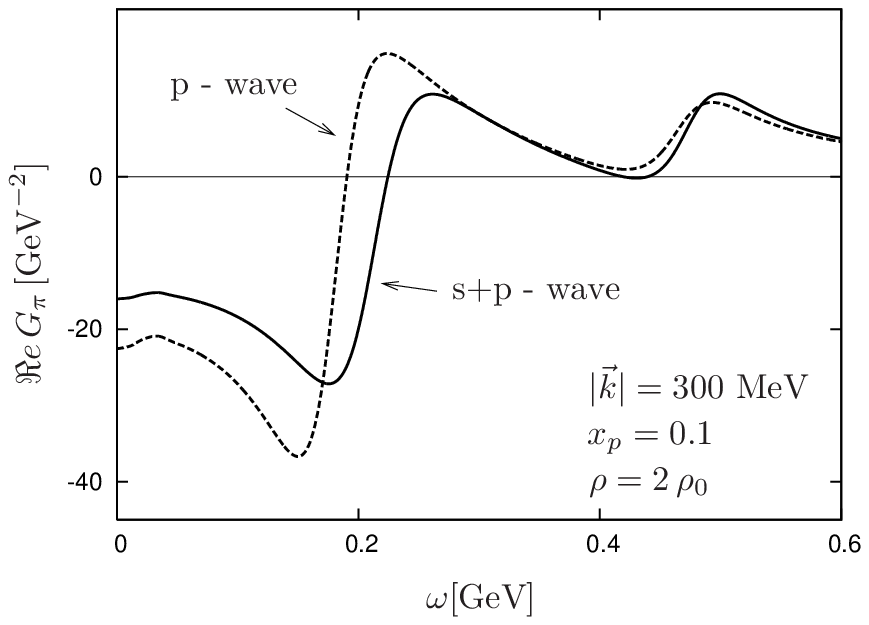}%}
  \end{center}
  \end{minipage}
 \begin{minipage}[h]{8.5cm}
 \begin{center}
 \includegraphics[width=8.5cm]{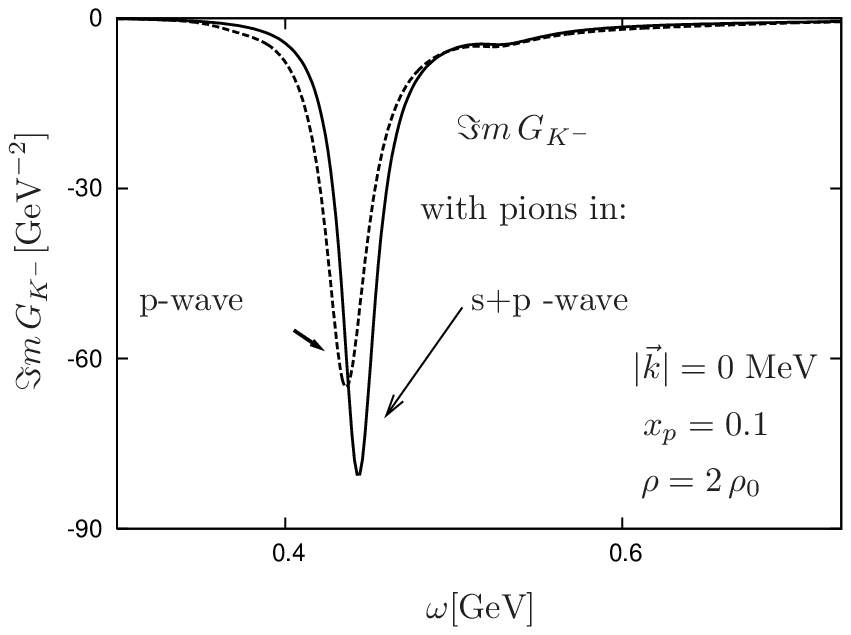}%}
 \end{center}
 \end{minipage}
 \begin{minipage}[h]{8.5cm}
 \begin{center}
 \includegraphics[width=8.5cm]{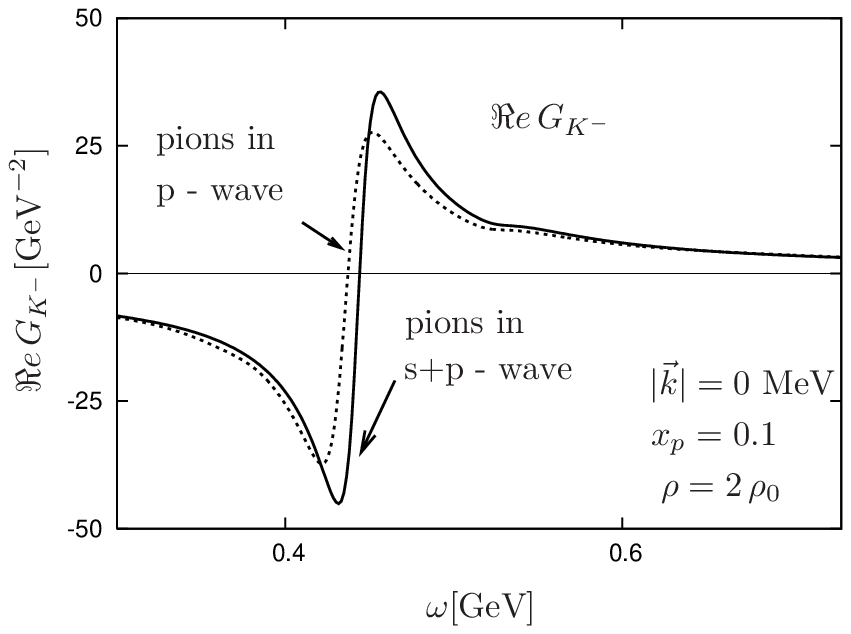}%}
 \end{center}
 \end{minipage}
 \caption[Propagators $G_{\pi}, \,G_{K^-}$, pions in $s + p$--wave]
  {
\small Upper panel:
  Imaginary (left) and real
  part (right) of the $\pi^-$ propagator as functions of energy for
  $|\vec k| = 300$~MeV. Solid lines: results
  including $s$--wave and $p$--wave contributions, dashed lines: 
  $p$--wave only.\\
Lower panel:
 Imaginary (left) and real
   part (right) of the $K^-$ propagator after the first iteration
   as functions of energy at $|\vec k| = 0$, 
   containing pions with including $s$--wave and $p$--wave 
   selfenergies (solid line) and  $p$--wave selfenergy only (dashed line).\\
All calculations in this figure correspond to total baryon density 
$\rho = 2 \rho_0$ and proton fraction $x_p = 0.1$.
}
 \label{figGpiGKm_ps}
 \end{figure}

It is therefore appropriate to check the influence of 
$s$--wave pion--nucleon interactions in the context of asymmetric nuclear
matter. To that end we employ the results of Ref.~\cite{pion_s_wave-Weise}
to calculate the pion selfenergy. 
The effect on the $\pi^-$--propagator  is illustrated in the upper panels of
Fig.~\ref{figGpiGKm_ps} for asymmetric 
nuclear matter with 10 \% protons at a total density of $2 \rho_0$. 
We see that the $\pi^-$ is pushed upwards in energy, i.e.,
the $s$--wave selfenergy is repulsive here. 

%%%%%%%%%%%%%%%%%%%%%%%%%%%%%%%%%%%%%%%%%%%%
%%%%%%%%%%%%%%%%%%%%%%%%%%%%%%%%%%%%%%%%%

However, we find again that the impact of the s--wave pion selfenergy on the 
antikaon propagator is rather weak. The lower panels of 
Fig.~\ref{figGpiGKm_ps} show the $K^-$ propagator after the
first iteration when the pions in the underlying $\pi Y$ loops are modified by
$p$ --wave and $s+p$ --wave interactions.
We infer from the figure that the $s$--wave $\pi$--selfenergy 
has some bearing on the antikaon 
propagator, but it is quite small. The zero of the real part of $G_{K^-}$ 
(the in--medium antikaon mass) is shifted 
 upwards by just a few MeV (right panel in Fig.~\ref{figGpiGKm_ps}).

 Thus the pion modifications in general have only minimal impact on the 
in--medium kaon.

%%==============================================================

\section{The kaon mass under neutron star conditions}
\label{chapKcondens}

As mentioned in the Introduction,
particular interest in the properties of antikaons in dense nuclear matter
is caused by the possibility of kaon condensation in neutron stars
\cite{KapNel,KapNel2}.

At not too high densities, neutron star matter consists of 
neutrons, protons, and electrons which are maintained in beta 
equilibrium through processes like 
$
      n \rightarrow p +  e^- + \bar\nu_e
$
and
$
      p + e^- \rightarrow n + \nu_e
$. 
For neutron stars older than a few minutes, the temperature is practically
zero on the nuclear scale, and the neutrino mean free path is larger than
the radius of the star, such that neutrinos can freely leave the system.
Hence, the electron chemical potential is equal to the difference between 
the chemical potentials of neutrons and protons, 
$\mu_e = \mu_n - \mu_p$.
Moreover, neutron star matter has to be electrically neutral. 
In the above case, this means that the densities of electrons and
protons must be equal, $\rho_e = \rho_p \equiv x_p \rho$.
Employing realistic equations of state,
one typically finds $x_p \simeq 0.1$ (see below). 
For the electron chemical potential, this roughly translates into
$\mu_e \simeq 150$~MeV~$(\rho/\rho_0)^{1/3}$.

With increasing density, various other weak processes may become 
energetically possible, leading to the occurrence of new particles.
In fact, according to the estimate above, already at $\rho = \rho_0$,
$\mu_e$ is well above the muon mass, and the conversion
$e^- \rightarrow \mu^- + \overline{\nu}_\mu + \nu_e$
should lead to a finite density of muons.  
At higher densities, strangeness--changing reactions should also occur. 
For instance, nucleons from the top of their Fermi seas could convert into 
hyperons.  

Similarly, electrons  and muons could convert into kaons,
\beq
e^- \rightarrow K^- + \nu_e~, \qquad \mu^- \rightarrow K^- + \nu_e~,
\label{electronKaon}
\eeq
once the electron ($=$ muon) chemical potential exceeds the in--medium mass 
of the $K^-$ \cite{Neutronstars-Wamb,KCondens1,KCondens3}.
This is the case we are interested in. 
Obviously, the conversion of electrons and muons to kaons would reduce the number
of fermions and thus their contribution to the pressure.
On the other hand, the kaons might form a condensate
at zero momentum that just provides a background of negative charge but, as a
condensate of bosons, does not exert a degeneracy pressure. 
In addition, processes like $n \rightarrow p + K^-$ would increase the   
proton fraction in the system. Since the isospin--asymmetry
part of the nuclear interaction is repulsive, this would cause a further
decrease of the pressure. Thus the formation of a condensate
could considerably soften the nuclear equation of state
\cite{LiLeeBrown,Neutronstars-Wamb}. 
In Ref.~\cite{BrownBethe}, this effect has been employed to predict
a maximum mass for neutron stars (better: nucleon stars) of about
$1.5$ solar masses and, as a consequence, the existence of a large
number of low--mass black holes in our galaxy. 

As pointed out above, a precondition for the conversion processes 
Eq.~(\ref{electronKaon}), is that 
the electron chemical potential must exceed the in--medium kaon mass, 
$m_{K^-}^*$. 
Strictly speaking, the relevant quantity is not the mass, but the
energy of the kaon. However, as we have seen in 
Sect.~\ref{pwavekaonselfenergy}, p--wave contributions to the kaon
self--energy are small, and therefore the lowest energies are found
at vanishing 3--momenta where they can be identified with the mass. 
Since the vacuum mass is much too large to be reached by $\mu_e$
at any realistic density, the question whether or not kaon condensation 
takes place in compact stars obviously depends crucially on how
fast $m_{K^-}^*$ drops as a function of density.
In the following we want to investigate this point with our model.

The mass of a given (quasi--) particle can be defined by the poles of the 
propagator.  
As explained above, we can restrict ourselves to the case of
vanishing 3--momentum. 
We also ignore the imaginary part of the selfenergy at this stage and
postpone its discussion to the end of this section. 
Then, the in--medium kaon mass is given by
\beq
{\Re} e\,G_{\Kbar}^{-1}(\omega= m_K^*,\vec{k}=0) = 
\left.\omega^2 -m_K^2- {\Re} e\, 
\Sigma_{\Kbar}(\omega,\vec{k}=0)\right|_{\omega= m_K^*} =0
\label{mstdef}
\eeq
Alternatively, we could define the mass as the maximum of the spectral
function. It turns out that the difference between these two definitions
is small for the cases of interest. 

As we have seen in Sect.~\ref{chapAsymm}, the kaon propagator depends on
both, total density $\rho$ and proton fraction $x_p$ of the surrounding 
medium. Hence, in order to proceed, we have to know $x_p$ as a function
of $\rho$. For neutron star matter this is in principle fixed by the
requirement of beta equilibrium and neutrality. However, to apply these
conditions, we need to know the equation of state which determines 
the relation between chemical potentials and densities of the involved
particles. 
As for the leptons, we can safely employ ideal gas relations.
On the other hand, this does not work for the hadrons where an ideal
gas equation of state yields unrealistically small proton fractions. 
Ultimately, of course, all hadrons should be described within a unified 
approach, where the in--medium pions and kaons are used to calculate
the dressed nucleons and vice versa. 
Lacking such a description at the present stage, we have consulted 
the literature to get some insight:
In Ref.~\cite{Neutronstars-Pand}, Akmal and collaborators
compare $x_p$ in neutral beta--stable matter resulting from 
a variety of nucleon--nucleon interactions, such as the Bonn, Urbana 
and Argonne models as well as the Nijmegen results. 
Typically, the values are of the order $x_p = 0.1$, with a general 
tendency that $x_p$ is rising with density. (For instance, taking 
their calculation based on the Argonne NN interaction $v_{18}$ and 
including boost corrections ($\delta_\nu$) and a parameterization of three 
nucleon forces ($UIX^*$), one finds $x_p = 0.06$ at $\rho = \rho_0$ and
$x_p = 0.14$ at $\rho = 5 \rho_0$, while the $v_{18}$ interaction alone
gives just $x_p = 0.095$ at $\rho = 5 \rho_0$.)
In view of these and other uncertainties and the fact that our results 
are not extremely sensitive to small variations of $x_p$, we take
a fixed value of $x_p = 0.1$ in our analysis. 

\begin{figure}[htp]
 \begin{center}
 \includegraphics[width=9.5cm]{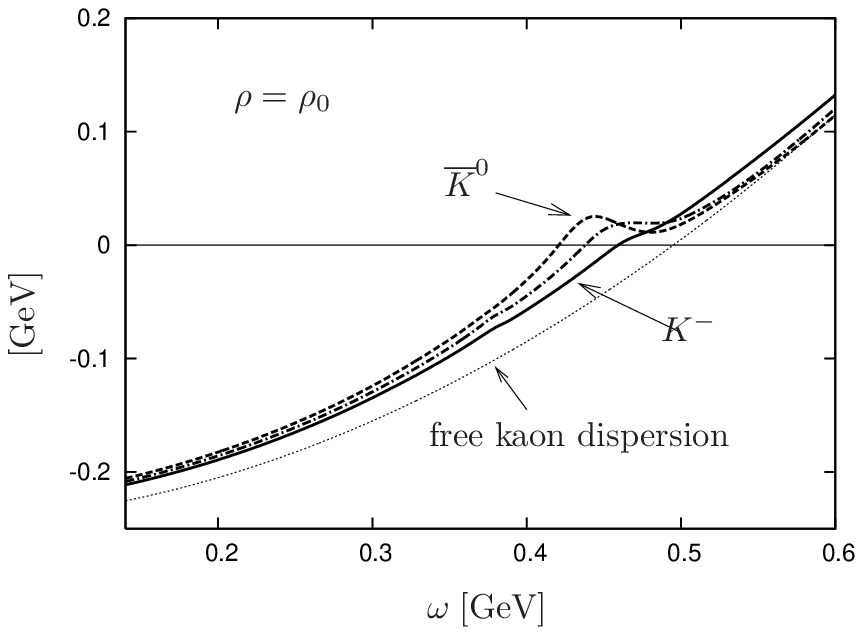}
 \includegraphics[width=9.5cm]{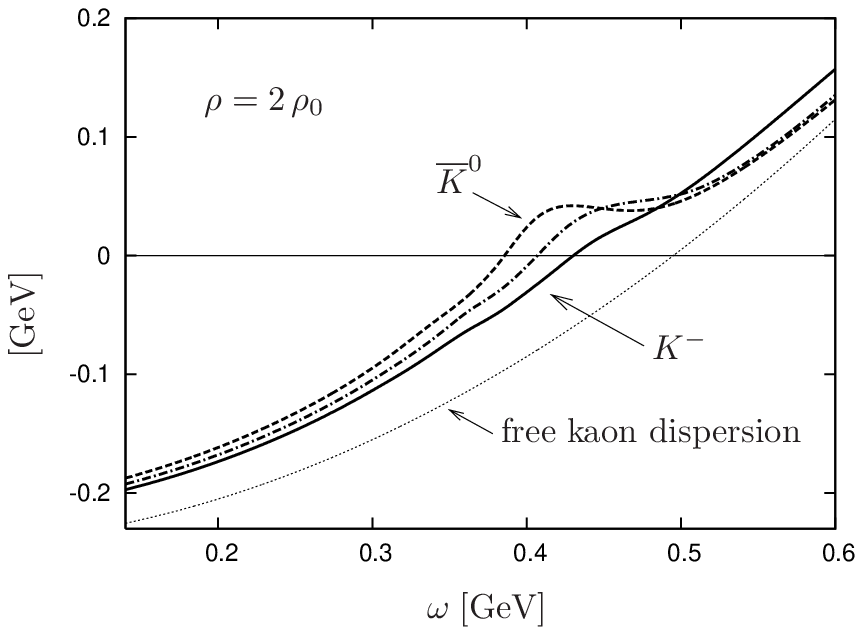}
 \includegraphics[width=9.5cm]{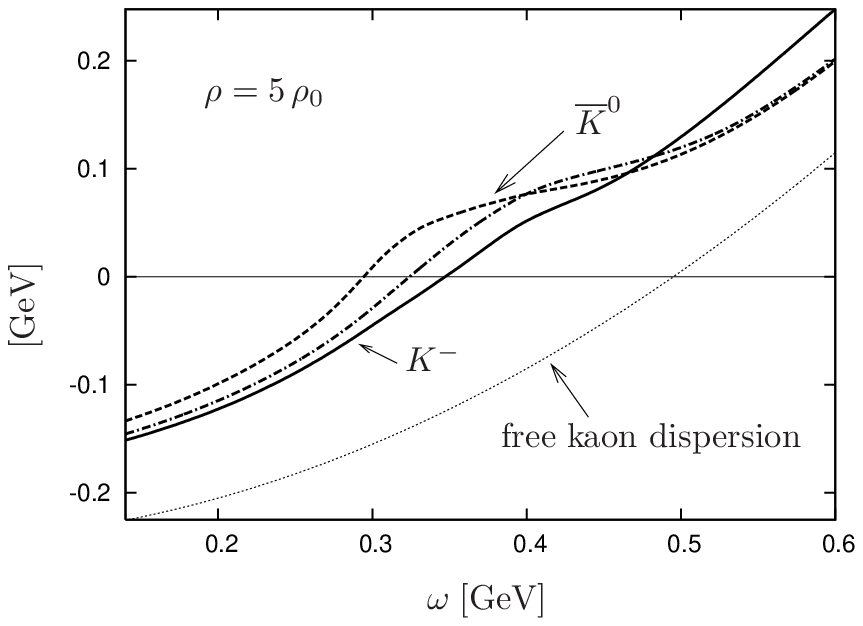}
 \caption[Antikaon mass at $u=5$]
{\small  Real part of the inverse kaon propagator 
         at vanishing momentum 3--momentum as a function of energy 
         for $x_p = 0.1$ and three different densities:
         $\rho = \rho_0$ (upper panel), $\rho = 2\rho_0$ (center), 
         and $\rho = 5\rho_0$ (lower panel).
         Solid line: $K^-$, dotted line: $\Kbar^0$,         
         dashed line: free kaon dispersion law,
         dashed--dotted line: $\Kbar$ for $x_p=0.5$}
 \label{KaonMass_10.90}
 \end{center}
\end{figure}

In Fig.~\ref{KaonMass_10.90} we show the real part of the inverse kaon
propagator $G_{\Kbar}^{-1}$  at vanishing momentum 3--momentum as a
function of energy. The three panels correspond to three different densities,
$\rho = \rho_0$, $\rho = 2\rho_0$, and $\rho = 5\rho_0$. 
For completeness both the $K^-$ (solid lines) and the $\Kbar^0$ 
(dashed lines) 
are shown, although only the $K^-$ mass is relevant for the question 
of kaon condensation.
For comparison, we also indicate the dispersion relation of the free
kaon (dotted lines).

%%%%%%%%%%%%%%%%%%%%%%%%%%%%%%%%%%%%%%%%%
%%%%%%%%%%%%%%%%%%%%%%%%%%%%%%%%%%%%%%%

The resulting effective kaon masses are listed in Table~\ref{tabmass}.  
As we have seen before, the neutral antikaons (which are irrelevant
in the context of kaon condensation in neutron stars) are more strongly 
affected by medium effects than the negative ones. 

For comparison with the literature we also list the effective mass for symmetric matter. 
At $\rho = \rho_0$ we find ${m}_{\Kbar}^{*\,symm}\!\! = 438$ MeV, corresponding to a optical 
potential depth of \mbox{$V_{opt}^{\Kbar} = \Sigma_{\Kbar N} / 2 {m}_{\Kbar}^{*} \approx -60$ MeV}. 
This is in good agreement with \cite{Galetal}, where \mbox{$V_{opt}^{\Kbar}\approx -55$ MeV} 
has been extracted from an analysis of kaonic atoms.

\begin{figure}[htp]
 \begin{center}
 \includegraphics[width=10cm]{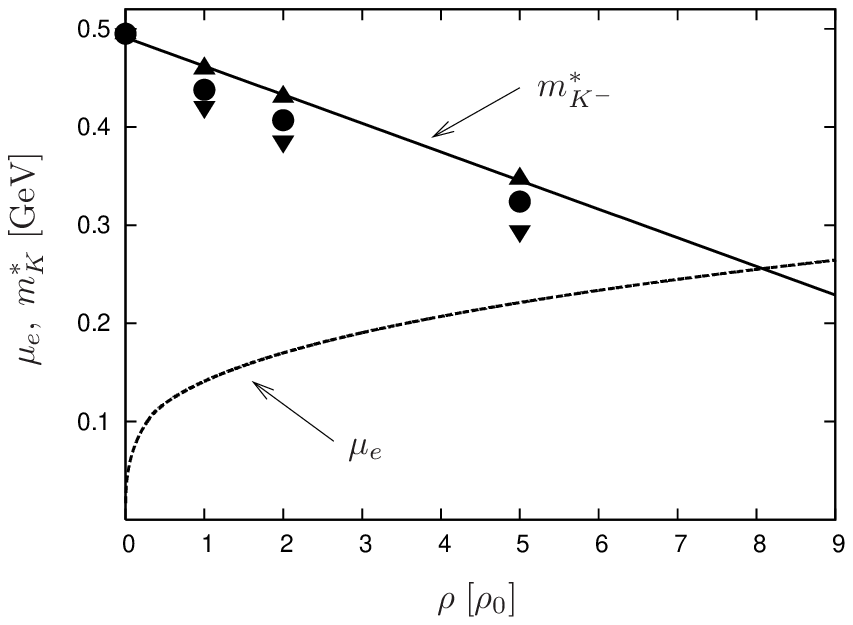}
 \caption
{\small In--medium $K^-$ mass $m_{K^-}^*$ ($\blacktriangle$) and electron 
        chemical potential $\mu_e$ (dashed line) versus baryon 
        density at $x_p=0.1$. The solid line represents a linear fit through the
        calculated points. Also indicated are the $\KObar$ mass 
        ($\blacktriangledown$) and the in--medium mass for the symmetric case ($\mathbf{\bullet}$).}
 \label{masses}
 \end{center}
\end{figure}

The masses are also plotted in Fig.~\ref{masses} as functions 
of $\rho$. The $K^-$ masses are indicated by $\blacktriangle$ symbols. 
We find that the numerical results are well described by a linear fit
\beq
    m_{K^-}^* (\rho) \;=\; 
    m_{K^-}(\rho= 0) \left( 1 \,-\, c \,\frac{\rho}{\rho_0} \right) 
\eeq
with $c = 0.06$ (solid line). 

%%%%%%%%%%%%%%%%%%%%%%%%%%%%%%%%%%%%%%%%%
\begin{table}[b!]
\begin{center}
\begin{tabular}{c c c c c}
\hline
&&&\\[-3mm]
\quad $\rho/\rho_0$ \quad & \quad $\mu_e$ [MeV] \quad & 
\quad ${m}_{K^-}^*$ [MeV] \quad & \quad ${m}_{\KObar}^*$ [MeV] \quad
& \quad ${m}_{\Kbar}^{*\,symm}$ [MeV] \quad
\\[2mm]
\hline
&&&\\[-3mm]
1 & 141 & 459 & 420 & 438 \\
2 & 170 & 430 & 385 & 407 \\
5 & 221 & 347 & 294 & 324 \\
\hline
\end{tabular}
\end{center}
\caption{\small Total baryon density and corresponding values of 
                the electron chemical potential and the in--medium
                masses of $K^-$ and $\KObar$ for $x_p = 0.1$. 
                Rightmost column: in--medium mass for $x_p = 0.5$}
\label{tabmass}
\end{table}  
%%%%%%%%%%%%%%%%%%%%%%%%%%%%%%%%%%%%%%%%%

%%%%%%%%%%%%%%%%%%%%%%%%%%%%%%%%%%%%%%%%%%
%%%%%%%%%%%%%%%%%%%%%%%%%%%%%%%%%%%%%%%%%

To decide whether the mass drop is sufficient to trigger kaon condensation
we need to know the corresponding values of the electron chemical potential
$\mu_e$. To that end, we consider a gas of non--interacting electrons and 
muons and determine $\mu_e$ by requiring charge neutrality together with
the protons in the system. The result is indicated by the dashed line in
Fig~\ref{masses}. The corresponding numbers are also listed in 
Table~\ref{tabmass}.

Obviously, $m_{K^-}^*$ is well above $\mu_e$, even at five times nuclear 
matter density. 
Even though the knowledge of nuclear interactions and the hadronic equation
of state under neutron star conditions is not very good, 
the variations in $x_p$ and $\mu_{e}$ caused by the uncertainties in 
present day calculations are not large enough to support the condensation 
scenario below $5 \rho_0$.
Extrapolating our results to higher densities, we predict the onset
of $K^-$ condensation at about $8 \rho_0$.
However, it is very unlikely that our model can be trusted up to
such high densities. In the hadronic sector there are several competing 
channels which probably open earlier and which we have neglected so
far. These could be, for instance, negative hyperons or pions.
Note that each additional negative source decelerates the rise of 
$\mu_e$ and therefore shifts the crossing point with the kaon mass to
higher densities.
Eventually, a phase transition to quark matter should take place. 
Hence, although a more detailed investigation of all these possibilities
is certainly necessary, we conclude that the occurrence of kaon 
condensation in neutron stars is rather unlikely. 

Finally, we should recall that in our determination of the 
in--medium kaon ``mass'' (cf. Eq.~(\ref{mstdef})), we have neglected the
imaginary part of the selfenergy. 
Therefore, one might raise the question whether the consideration of the 
kaon width could change our conclusions. 
However, this seems not to be the case:
As one can see in Fig.~\ref{ImGKm0_u_10.90_k0}, even the 
low--energy tails of the spectral functions are far away from the 
values of $\mu_e$ given in Table~\ref{tabmass}.

In this context
we should emphasize that our present approach is not
suited to describe the phase transition to a kaon condensed phase itself. 
If we further increase the density, eventually part of the spectral 
function will move below $\mu_e$ while another part still lies above,
and it would be unclear whether or not the conversion process 
Eq.~(\ref{electronKaon}) takes place\footnote{
Strictly speaking, this problem exists already at lower densities
in our model: If we cut the kaon selfenergy in Fig.~\ref{overview_fig}, 
we see that there is a contribution corresponding to the decay
$K^- \rightarrow \pi^0 \Lambda p^{-1}$, where $p^{-1}$ denotes a hole
state in the Fermi sea of the protons. 
Moreover, the pion can further decay into a nucleon--hole state.
Therefore the imaginary part of the kaon propagator
opens at an energy approximately equal to the $\Lambda$--proton 
mass difference, where we have neglected the Fermi energy of the 
protons.
This means that part of the kaon spectral function is formally 
below the electron chemical potential as soon as the latter exceeds this 
value which is the case at about $2 \rho_0$.}.
This is an artifact of our model where we calculate the modified 
kaon propagator, leaving the ground state of the system, i.e., the 
Fermi sea of nucleons, unchanged. 
It is clear that, once kaon condensation sets in, the kaons with vanishing
3--momentum are part of the ground state, and the width of the propagator
has to vanish at $\omega = \mu_e$. 
Hence, our present approach is only justified as long as the modifications
of the nucleonic ground state are small in comparison with the kaon sector,
i.e., as long as we are far away from the phase transition. 
In this sense, at $\rho = 5\rho_0$ we are still on the safe side.

%%==============================================================
\section{Summary}
\label{cSummary}

We have investigated the properties of antikaons in dense nuclear
matter, focusing on the regime of zero temperature and high densities. 
Taking the most important $KN$ vertices derived in $\chi PT$: the
Weinberg--Tomozawa term and the $\sigma$ terms,
we have constructed the $T$--matrix for meson--baryon scattering within
a coupled--channel approach, 
thereby dynamically generating the $\Lambda(1405)$ resonance.
Regularization of the scattering amplitudes was accomplished by means 
of twice subtracted dispersion relations. 
The values of the subtraction constants have been
fixed by a fit to the experimental scattering lengths. 

Having developed a satisfactory description of the scattering process
in vacuum, the calculations have been extended to the dense medium.
The medium was treated in the nucleon gas approximation, i.e., 
assuming a free Fermi gas of protons and neutrons. 
This leads to a modified $T$--matrix through Pauli blocking 
the occupied nucleon states.  
On the other hand, closing a nucleon (hole) loop in the $T$--matrix, 
we obtained the in--medium self--energy of the antikaon. 
In turn, the resulting kaon propagator was used to (re--) calculate the 
$T$--matrix.  
This procedure was iterated until a selfconsistent result was achieved.
Convergence was reached after some four to five iterations.
The actual implementation of the selfconsistency scheme requires the 
determination of the scattering amplitudes and the antikaon propagator 
for the entire energy--momentum plane.
A suitable method to implement this requirement was developed. 

We first applied the model to isospin symmetric nuclear matter.
We found that the $\Lambda(1405)$ resonance is moved to higher energies 
and becomes strongly broadened. The antikaon also receives a finite width 
and is mostly shifted to lower energies.  
Its spectral function is found to be strongly momentum dependent, 
emphasizing  the need for a calculation that correctly incorporates 
the finite 3--momentum  of the scattering process and the in--medium antikaon. 
The spectral function at zero momentum is broad but the antikaon mass 
still seems to be well defined. At finite but small momenta there are 
two branches that stem from the in--medium excitations at lower energies 
and the antikaon pole at $\sqrt{s} \approx 500$ MeV. 
At higher momenta the antikaon dispersion relation approaches its free form.

Besides the momentum dependence, we found the selfconsistent treatment of
$T$--matrix and kaon propagator to have the largest effect on the results,
while other features, like the in--medium modification of the pions and
$p$--wave $KN$--interactions were found to be less important.

We then applied our model to isospin--asymmetric matter as found inside 
neutron stars. 
Although the general dependence on density and momentum is similar to that
seen for symmetric matter, the most important difference is that, in 
asymmetric matter, $K^-$  and $\KObar$ develop distinct spectral functions.
This could be traced back to the Weinberg--Tomozawa term which is strongest
in the $\KObar n$ and $K^- p$ channels.

Finally, we investigated the possibility of $K^-$ condensation in 
neutron stars. 
Condensation should set in when the $K^-$ mass falls below the electron 
chemical potential in electrically neutral beta equilibrated matter. 
Neglecting the width of the $K^-$, we found that the mass can be 
described rather well by a linear fit, $m_{K^-}^* (\rho) = 
m_{K^-}(\rho= 0) ( 1 - 0.06 \rho/\rho_0)$.
This drop is too slow to support kaon condensation for densities 
up to at least $5 \rho_0$. 

%%=============================================================
\section*{Acknowledgments}

We are grateful to M.~Lutz for illuminating discussions and the possibility of comparing our
vacuum amplitudes with his results.
We thank M.~Urban and A.~Wirzba for expert advice on $\chi PT$ and technical questions, 
and A.~Gal. for instructive comments.
Th.~R. acknowledges support by COSY contract 41315274 (COSY--039).
%%=============================================================

\end{document}